\documentclass[
onecolumn,
nofootinbib,
preprint,
tightenlines,
amsmath,amssymb,amsthm,
]{revtex4-2}
\usepackage{xcolor}
\usepackage[utf8]{inputenc}
\usepackage{mathtools}
\usepackage{amsmath}
\usepackage{amsthm}
\usepackage{amsbsy}
\usepackage{amssymb}
\usepackage{amscd}
\usepackage{amsfonts}
\usepackage{bm}
\usepackage{graphics}
\usepackage{graphicx}
\usepackage[plain]{fancyref}
\usepackage{soul}
\usepackage{hyperref}
\setcitestyle{numbers,sort&compress}
\usepackage{verbatim}
\usepackage[margin=25mm]{geometry}
\usepackage{paralist}
\usepackage{wasysym}
\usepackage{lineno}

\graphicspath{ {figs/} }

\newcommand*{\fancyrefapplabelprefix}{app}
\frefformat{plain}{\fancyrefapplabelprefix}{appendix~#1}
\Frefformat{plain}{\fancyrefapplabelprefix}{Appendix~#1}
\frefformat{main}{\fancyrefapplabelprefix}{appendix~#1}
\Frefformat{main}{\fancyrefapplabelprefix}{Appendix~#1}

\newcommand{\partialx}[2]{\frac{\partial #1}{\partial #2}}
\newcommand{\partialN}[3]{\frac{\partial^{#3} #1}{\partial #2^{#3}}}
\newcommand{\partialInline}[2]{\partial #1 / \partial #2}

\newcommand{\diverge}[1]{\nabla \cdot #1}

\newcommand{\takegrad}[1]{\nabla #1}

\newcommand{\takeLapl}[1]{\nabla^2 #1}

\newcommand{\df}[1]{\text{d}#1} 
\newcommand{\Df}[1]{\mathcal{D}#1} 

\newcommand{\Reals}{\mathbb{R}}

\newcommand{\kB}{k} 
\newcommand{\T}{T} 

\newcommand{\avg}[1]{\left< #1 \right>}
\newcommand{\generic}{\Box}

\newcommand{\orderOf}[1]{\mathcal{O}\left(#1\right)}

\newcommand{\smallpar}{\epsilon}

\newcommand{\U}{u}
\newcommand{\V}{v}
\newcommand{\creaseAngle}{\theta}
\newcommand{\creaseAngleZero}{\creaseAngle_0}
\newcommand{\creaseAngleStar}{\creaseAngle^*}
\newcommand{\creaseAnglePM}{\creaseAngle^{\pm}}
\newcommand{\creaseAngleP}{\creaseAngle^+}
\newcommand{\creaseAngleL}{\creaseAngle^-}
\newcommand{\creaseAngleClosed}{\creaseAngle^c}
\newcommand{\creaseAngleFoldedMin}{\creaseAngle_{f}}
\newcommand{\creaseAngleBarrier}{\creaseAngleStar}
\newcommand{\creaseAngleU}{\creaseAngle_u}
\newcommand{\vdWToEnt}{R}

\newcommand{\D}{d}
\newcommand{\DO}{\D_0}
\newcommand{\DAvg}{\bar{\D}}
\newcommand{\len}{\ell}
\newcommand{\Lu}{\len_{\U}}
\newcommand{\Lv}{\len_{\V}}
\newcommand{\area}{A}
\newcommand{\lt}{a}
\newcommand{\h}{h}
\newcommand{\hCons}{\tilde{h}}
\newcommand{\domain}{\Omega}

\newcommand{\Ham}{\mathcal{H}}
\newcommand{\Hcrease}{\Ham_\creaseAngle}
\newcommand{\kcrease}{\kappa_\creaseAngle}
\newcommand{\sPot}{\sigma}
\newcommand{\HvdW}{\Ham_\sPot}
\newcommand{\Hbend}{\Ham_b}
\newcommand{\HGauss}{\Ham_G}
\newcommand{\kbend}{\kappa}
\newcommand{\kGauss}{\kappa_G}
\newcommand{\vPot}{\nu}

\newcommand{\meanCurv}{H}
\newcommand{\GaussCurv}{K}
\newcommand{\CvdW}{C_W}
\newcommand{\pPot}{\gamma}

\newcommand{\intOverSurf}[1]{\int_0^{\Lu} \df{\U} \int_0^{\Lv} \df{\V} \: #1}
\newcommand{\partFunc}{\mathcal{Z}}
\newcommand{\freeEnergy}{\mathcal{A}}

\newcommand{\Nats}{\mathbb{N}}
\newcommand{\qm}{q_m}
\newcommand{\pn}{p_n}
\newcommand{\Amn}{A_{mn}}
\newcommand{\Bmn}{B_{mn}}
\newcommand{\Cmn}{C_{mn}}
\newcommand{\Dmn}{D_{mn}}
\newcommand{\modeNums}{\mathbb{M}}
\newcommand{\cqm}{\cos \left(\qm \U\right)}
\newcommand{\sqm}{\sin \left(\qm \U\right)}
\newcommand{\cpn}{\cos \left(\pn \V\right)}
\newcommand{\spn}{\sin \left(\pn \V\right)}

\newcommand{\prob}{\rho}
\newcommand{\stericFactor}{\mu}
\newcommand{\freeEnergyConst}{\freeEnergy_0}
\newcommand{\diracFunctional}[1]{\delta \left[#1\right]}
\newcommand{\diracDelta}[1]{\delta_{#1}}
\newcommand{\waveFunc}{\omega}
\newcommand{\Vk}{V_k}
\newcommand{\Wk}{W_k}
\newcommand{\Vn}{V_n}
\newcommand{\Wn}{W_n}
\newcommand{\Hmn}{H_{mn}}
\newcommand{\freeEnergyH}{\freeEnergy_{\h}}

\newcommand{\torqueH}{\mathcal{T}_{\h}}
\newcommand{\torqueSto}{\eta}
\newcommand{\tv}{t}
\newcommand{\Imom}{I}
\newcommand{\drag}{\gamma}
\newcommand{\currentDensity}{j}
\newcommand{\rate}{K}
\newcommand{\nbr}[1]{\mathcal{N}\left(#1\right)}
\newcommand{\probEq}{\prob^{\mathrm{eq}}}
\newcommand{\currentDensityEq}{\currentDensity^{\mathrm{st}}}
\newcommand{\dA}{\Delta \freeEnergy}
\newcommand{\rateTST}{\rate_{\mathrm{TST}}}
\newcommand{\rateKr}{\rate_{\mathrm{Kr}}}
\newcommand{\density}{\sigma}

\newcommand{\aspectRatio}{\lambda}

\newcommand{\eV}{\textnormal{eV}}
\newcommand{\eVSp}{\textnormal{eV }}
\newcommand{\angs}{\textnormal{\r{A}}}
\newcommand{\angsSp}{\textnormal{\r{A} }}
\makeatletter
\newcommand*{\gnuplotinput}[2][1.0]{%
  \begingroup
  \let\@gnplt@input@includegraphics=\includegraphics
  \def\includegraphics##1{\@gnplt@input@includegraphics[scale=#1]{#2}}%
  \let\@gnplt@input@picture=\picture
  \def\picture{\unitlength=#1\unitlength\relax\@gnplt@input@picture}%
  \input{#2}%
  \endgroup
}
\makeatother

\theoremstyle{remark}

\newcommand*{\fancyrefproplabelprefix}{prop}
\frefformat{plain}{\fancyrefproplabelprefix}{proposition~#1}
\Frefformat{plain}{\fancyrefproplabelprefix}{Proposition~#1}
\frefformat{main}{\fancyrefproplabelprefix}{proposition~#1}
\Frefformat{main}{\fancyrefproplabelprefix}{Proposition~#1}

\newcommand*{\fancyreflemlabelprefix}{lem}
\frefformat{plain}{\fancyreflemlabelprefix}{lemma~#1}
\Frefformat{plain}{\fancyreflemlabelprefix}{Lemma~#1}
\frefformat{main}{\fancyreflemlabelprefix}{lemma~#1}
\Frefformat{main}{\fancyreflemlabelprefix}{Lemma~#1}

\newcommand{\capTitle}[1]{\emph{#1}}

\renewcommand{\hl}[1]{#1}
{}%
{}

\begin{document}

\title{
	Thermal fluctuations (eventually) unfold nanoscale origami
}


\author{Matthew Grasinger}
\email{matthew.grasinger.1@us.af.mil}
\affiliation{Materials and Manufacturing Directorate, Air Force Research Laboratory}

\author{Pradeep Sharma}
\email{psharma@uh.edu}
\affiliation{University of Houston, Department of Mechanical Engineering}
\affiliation{University of Houston, Department of Physics}

\preprint{To appear in Journal of the Mechanics and Physics of Solids, doi: \href{https://doi.org/10.1016/j.jmps.2023.105527}{10.1016/j.jmps.2023.105527}.}


\begin{abstract}
  Origami is a scale invariant paradigm for morphing robotics, deployable structures (e.g. satellites, disaster relief shelters, medical stents), and metamaterials with tunable thermal, mechanical, or electromagnetic properties. 
  There has been a resurgence of interest in using origami principles--along with $2$D materials or DNA--to design a wide array of nanoscale devices. 
In this work, we take cognizance of the fact that small-scale devices are vulnerable to entropic thermal fluctuations and thus a foundational question underlying small-scale origami pertains to its stability, i.e. the origami structure's propensity to ``unfold" due to thermal fluctuations and the rate at which the unfolding will ensue. To properly understand the behavior of these origami-based nanodevices, we must simultaneously consider the geometric mechanics of origami along with the interplay between thermal fluctuations, entropic repulsive forces, van der Waals attraction, and other molecular-scale phenomena. In this work, to elucidate the rich behavior underpinning the evolution of an origami device at the nanoscale, we develop a \emph{minimal} statistical mechanics model of folded nanoscale sheets. We use the model to investigate \begin{inparaenum}[(1)] \item the thermodynamic multistability of nanoscale origami structures and \item the rate at which thermal fluctuations drive its unfolding--that is, its temporal stability. \end{inparaenum} We identify, for the first time, an \emph{entropic} torque that is a critical driving force for the unfolding process. Both the thermodynamic multistability and temporal stability have a nontrivial dependence on the origami’s bending stiffness, the radii of curvature of its creases, the ambient temperature, its thickness, and its interfacial energy (between folded layers). Specifically, for graphene, we show that there is a critical side length below which it can no longer be folded with stability; similarly, there exists a critical crease diameter, membrane thickness (e.g. for multilayer graphene), and temperature above which a crease cannot be stably folded. To investigate the rate of thermally driven unfolding, we extend Kramers' escape rate theory to cases where the minima of the energy well occurs at a boundary. Rates of unfolding are found to span from effectively zero to instantaneous, and there is a clear interplay between temperature, geometry, and mechanical properties on the unfolding rate.
\end{abstract}

\maketitle

\section{Introduction} \label{sec:intro}

Origami engineering is rapidly emerging as an important modality to create novel classes of metamaterials. Such ``materials'' may display properties that are not found in natural materials, offer prospects for tuning thermal~\cite{boatti2017origami}, mechanical~\cite{liu2022triclinic,miyazawa2021heterogeneous,misseroni2022experimental,zhai2020situ,silverberg2014using,schenk2013geometry,liu2018topological,grasinger2022multistability,pratapa2019geometric,brunck2016elastic} or electromagnetic~\cite{sessions2019origami} response and are anticipated to find applications in fields ranging from robotics~\cite{novelino2020untethered,wu2021stretchable,fang2017origami}, mechanologic \cite{treml2018origami}, deployable civil engineering structures to cloaking devices. Extensive literature already exists in this context that runs the gamut from fundamental design rules, mathematical theorems~\cite{kawasaki1991relation,lang2017twists,demaine2007geometric,hull2002modelling}, fabrication paradigms, and discipline-specific applications. \\

Our focus in this work is origami behavior at the small scale. The literature is rife with proposals for myriad applications of molecular origami such as targeted drug delivery \cite{jiang2019rationally,udomprasert2017dna,weiden2021dna,ge2020dna,chandrasekaran2016beyond}, nanorobotics \cite{liu2023light}, energy harvesting at the nanoscale, optics \cite{cho2011nanoscale}, and molecular origami nanocomposites~\cite{zhao2022enhanced,zhao2021significantly}--which combine the novel behaviors of origami-based metamaterials (e.g. auxetic materials~\cite{pratapa2019geometric}, tunable stiffness \cite{silverberg2014using} and thermal expansion \cite{boatti2017origami}, phononic metamaterials, space-time metamaterials) with the high strength-to-weight ratios of composites. \\

While there are still several open issues in modeling of large-scale origami\footnote{With our focus on nanoscale, even micron-scale devices would be classified as ``large-scale" in our terminology}, the field is developing rapidly. In contrast, both experimentally and theoretically, nanoscale origami is relatively at its infancy. Aside from the challenges of fabrication, the key difference between small-scale origami and its counterpart at macroscopic length scale is due to the anticipated influence of thermal fluctuations. A good analogy may be drawn by comparing a conventional micron-scale thin film to a graphene sheet. While entropic effects are irrelevant for the thin film, their influence on graphene strongly dictates both its mechanical and electronic behavior \cite{ahmadpoor2017thermal,fasolino2007intrinsic,thibado2020fluctuation, de2008periodically}. Thus, the central question may be posed as follows: how stable is a small-scale origami structure to thermal fluctuations? To address this, in our work, we consider the simplest possible embodiment that may be regarded as a proxy for the origami structure i.e. a folded sheet with a crease. Even this simple system is expected to show very rich behavior. This may be understood easily by a few examples. Consider two elastic sheets in thermal equilibrium. Both will undergo thermal fluctuations. The strength of the fluctuation is dictated by the energy cost parametrized by the elasticity of the sheets. Given that elastic energy cost of a fluctuation of a thin sheet is much lower than that of a thicker one, we can easily understand why 2D materials like graphene or biological cell membranes fluctuate noticeably while such effects can be neglected in micron or larger thickness films. If the thin sheets are far enough from each other then they will fluctuate independently. As the sheets are brought closer together, each sheet will hinder  the allowed fluctuations of the other. This hindrance reduces the entropic freedom and thus increases the free-energy of the 2-sheet system. In other words, this steric hindrance results in a repulsive force of entropic origin that is now well-known to vary as $1/d^3$, where $d$ is separation between parallel thin sheets. This result is well-recognized in the biophysics community and has been extensively studied c.f.\cite{helfrich1978steric,helfrich1984undulations, hanlumyuang2014revisiting,freund2013entropic,lu2015effective, wennerstrom2014undulation, mozaffari2021flexoelectricity, liang2018method} as well in applications related to 2D materials \cite{ahmadpoor2022entropic, zhu2022thermal} and even polycrystals \cite{chen2015entropic}. The entropic repulsive force in that context has wide-ranging implications from cell fusion, endocytosis, self-assembly among others. In our context, a folded sheet that serves as a proxy for origami structure is even more complex. First, the two halves of the folded sheet are likely to be in close proximity so the repulsive force is expected to be quite significant; second, the folded edge provides a strong constraint on the problem and thus the fluctuations are not homogeneous but vary along the length of the sheet. In other words, the crease at the fold provides for a hinge around which the folded structure could potentially unfold with the entropic repulsive force that varies spatially. The close proximity of the two halves of a folded sheet also imply the strong (likely) role of a competing attractive force: the van der Waals attraction.\\

We cite several recent interesting works that have modeled (largely numerical) some facet of origami or folding/unfolding of thin sheets e.g.  Mannatti et. al. \cite{mannattil2022thermal} examined a single-bar joint system, Rocklin et. al. \cite{rocklin2018folding} modeled folding mechanisms of thin sheets, Yong and Mahadevan \cite{yong2014statistical} performed non-equilibrium simulations to elucidate the shape transitions in plates, and Yang et. al. \cite{yang2021energetics} performed molecular dynamics simulations to look at folded graphene sheets. These works are valuable contributions but differ from ours in that we attempt to provide an analytical paradigm to understand the various mechanisms that control the unfolding of a folded sheet with a crease. In particular, our model approach provides a transparent framework that segregates and accounts for the interplay of the various forces that operate at small-scale e.g. entropic repulsion due to thermal fluctuations, van der Waals attraction and nonlinear geometrical mechanics. Finally, an important element of our treatment of the statistical mechanics problem pertains to the way we resolve the non-equilibrium nature of the unfolding process. We recognize that the unfolding process may be interpreted in terms of the deformation of the sheet by slow modes and fast modes (e.g. thermal undulations). There is a distinct separation of time scales and we take advantage of this by first solving the equilibrium statistical mechanics of the folded sheet at a fixed unfolded angle. Subsequently, we model the non-equilibrium evolution of the slow mode (e.g. changes in the folding angle) via the Langevin equation in the diffusion regime (i.e. overdamped limit) where the previous, quasi-equilibrium results show up as a thermodynamic driving force.  \\


In light of the context pertaining to small-scale origami discussed in the preceding paragraphs, in this paper, we:

\begin{itemize}
    \item {Create a minimal model of a folded (creased) sheet that characterizes the interplay of entropic repulsion, van der Waals attraction, and nonlinear origami mechanics.}
    \item {Present a new approach for modeling repulsion of membranes at an angle with respect to each other and derive results for fluctuating membranes with hinged boundary conditions.}
    \item {Examine the thermodynamic stability of small-scale origami at finite temperature both in closed-form and via numerical examples.}
    	\item {Provide insights into the multistability of the origami problem in relation to temperature, radii of curvature of the fold crease, planar dimensions, geometry among others.}
    \item {Elucidate temporal stability of small-scale origami by exploiting transition state theory (TST) and a modified Kramers' escape rate approach. We remark that Kramers' escape rate is specific to energy landscapes with interior local minima, and it cannot account for minima at a boundary where the gradient of the energy is nonzero. In this work, we extend Kramers' approach to consider local minima at a boundary.}
\end{itemize}

\section{Statistical mechanics of folded membranes} \label{sec:}

\begin{figure*}
	\centering
	\includegraphics[width=\linewidth]{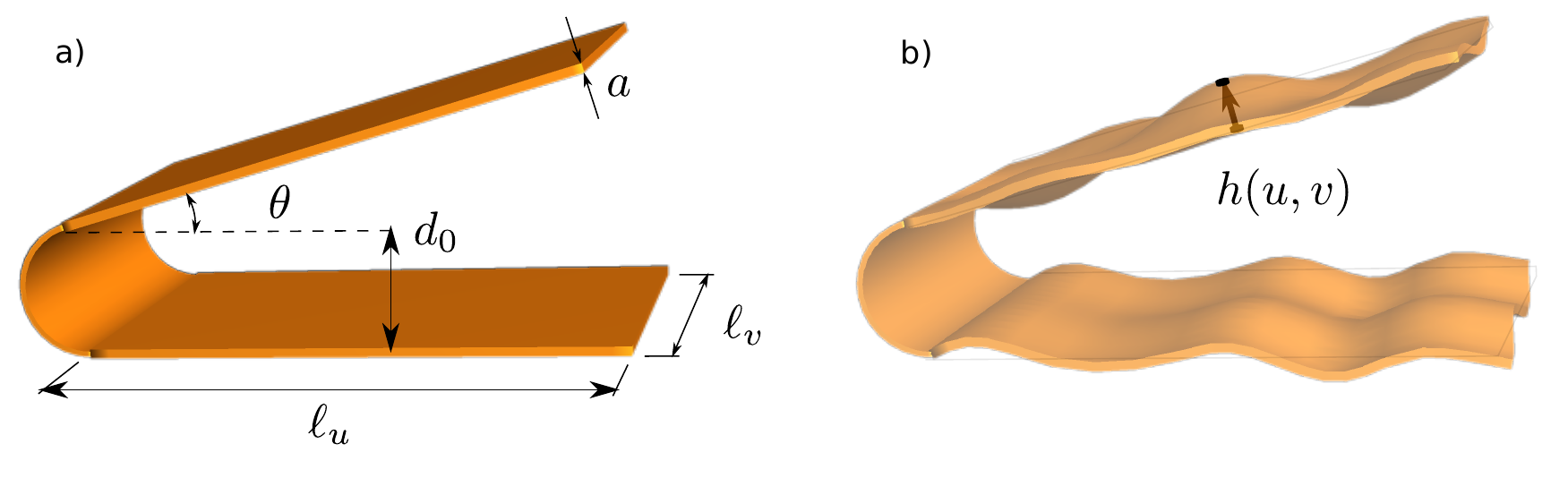}
	\caption{
		\capTitle{Molecular origami macrostate and snapshot of a microstate.}
		a) Thermodynamic state variables of the molecular origami system of interest: length of the sides of the top and bottom membranes, $\Lu$ and $\Lv$, angle of elevation of the top crease, $\creaseAngle$, distance between the top and bottom membranes at $\creaseAngle = 0$, $\DO$, and membrane thickness, $\lt$.
		b) Example microstate with thermal undulations in the top and bottom membranes, $\h\left(\U, \V\right)$ where $\h$ is the displacement out of the plane of the membrane.
	}
	\label{fig:folded-states}
\end{figure*}

We consider a flat molecular sheet folded in half about a single crease, as shown in \fref{fig:folded-states}.a.
We call the angle between the top and bottom folded membranes the ``crease angle'', $\creaseAngle$; for example, $\creaseAngle = 0$ when the two membranes are parallel and $\creaseAngle = \pi$ when the sheet is flat.
The crease angle is assumed to have negligible variance--except at timescales several orders of magnitude larger than the timescale of atomic vibrations.
As a result, the crease angle can be used to describe the equilibrium, macroscopic state (i.e. ``macrostate'' in the thermodynamic sense) of the molecular origami.
Assume for now that $\creaseAngle \leq \pi / 2$.
The crease has a finite radius such that, when $\creaseAngle = 0$, there is a distance $\DO$ between the two membranes.
The two membranes have thickness $\lt$, planar dimensions $\Lu$ and $\Lv$, and material points within the top and bottom are parameterized by $\left(\U, \V\right) \in \domain \cong \left[0, \Lu\right] \times \left[0, \Lv\right]$ and $\left(\U', \V'\right) \in \domain' \cong \left[0, \Lu\right] \times \left[0, \Lv\right]$, respectively.

Molecules within the two membranes fluctuate about their mean positions at finite temperature--leading to deformations, or \emph{undulations}, of the top and bottom membranes.
To characterize the deformed microstates of the sheets, we draw inspiration from past work on the statistical mechanics of biological membranes and utilize the Monge gauge~\cite{mozaffari2021flexoelectricity,hanlumyuang2014revisiting,kulkarni2023fluctuations}.
The Monge gauge, $\h : \domain \cup \domain' \mapsto \Reals$, is a scalar displacement field where the displacement is assumed orthogonal to the plane of the folded, but otherwise flat, top and bottom membrane\footnote{Any undulations within the crease itself are not explicitly considered because \begin{inparaenum}[1)] \item the crease is a curved section of membrane and, as a consequence, has a higher apparent bending stiffness orthogonal to the line of the crease, and \item the area of the crease is considered small relative to the rest of the folded membrane such that the free energy contribution of its undulations are negligible. \end{inparaenum}}.
Although limiting our attention to deformations orthogonal to the undeformed sheet may seem overly restrictive, it represents a good approximation for allowable deformations because \begin{inparaenum}[1)] \item the molecular sheets of interest are effectively rigid to in-plane, stretching deformations relative to bending and \item because out-of-plane deformations are isometric when the deformations are small enough~\cite{chen2018branches,zhou2023low}. \end{inparaenum}

\subsection{Energetics} Next we characterize the kinematics and associated energetics of a microstate of the folded molecular sheet.
Let
\begin{equation} \label{eq:dist}
	\D\left(\U, \V\right) = \DO + \U \tan \creaseAngle + \h\left(\U, \V\right) \cos \creaseAngle,
\end{equation} be 
the distance between $\left(\U, \V\right)$ and the bottom membrane, as measured orthogonal from the top membrane. 
Further, let the combined spatial and thermodynamic average of the distance between the fluctuating membranes and the wall be $\DAvg$.
By assumption, the thermodynamic average of the undulations, $\avg{\h}$, vanishes so that
\begin{equation}
	\DAvg = \DO + \frac{\Lu}{2} \tan \creaseAngle.
\end{equation}
Then the energy of the folded molecular sheet consists of \begin{enumerate}[1)] 
	\item the bending energy of the crease which is idealized as a linear torsional spring, \begin{equation} \label{eq:creaseEnergy} 
		\Hcrease = \frac{\kcrease \Lv}{2} \left(\creaseAngle - \creaseAngleZero\right)^2, 
	\end{equation}
	where $\kcrease$ is the torsional stiffness per length of the crease.
	\item the van der Waals attraction between the membranes, 
        \begin{equation} \label{eq:vdW-approx}
		\HvdW \approx -\frac{\CvdW \area}{12 \pi} \left(\frac{1}{\DAvg^2} - \frac{2}{\left(\DAvg + \lt\right)^2} + \frac{1}{\left(\DAvg + 2\lt\right)^2}\right),
		\end{equation}
        where $\area = \Lu \Lv$ is the area of a folded membrane\hl{, and $\CvdW$ is the Hamaker constant, which characterizes the strength of the (body-body) van der Waals interaction ($\CvdW > 0$ corresponds to an attractive interaction, $\CvdW$ corresponds to repulsive)}.
        Equation \eqref{eq:vdW-approx} is an adaptation of a well-known result for the van der Waals potential between two parallel membranes~\cite{tadmor2001london,mozaffari2021flexoelectricity} where $\DAvg$ is used in place of ``distance'' in order to account for the change in average distance and screening of the potential that occur as a function of $\creaseAngle$.
        
	And,
	\item the bending energy of the membranes due to undulations. 
         Here, up to quadratic order, there are contributions from the mean curvature, $\Hbend = \kbend \meanCurv^2 / 2$, and the Gaussian curvature, $\HGauss = \kGauss \GaussCurv$~\cite{helfrich1973elastic} where
    $\kbend$ is the bending modulus, $\kGauss$ is the Gaussian modulus, $\meanCurv$ is the mean curvature, and $\GaussCurv$ is the Gaussian curvature.
    In terms of the Monge representation, the mean and Gaussian curvatures are
    \begin{subequations}
    \begin{align}
        \meanCurv &= \diverge{\left( \frac{\takegrad{\h}}{\sqrt{1 + \left|\takegrad{\h}\right|^2}} \right)} \approx \takeLapl{\h}, \\
        \GaussCurv &= \frac{\det\left(\takegrad{\takegrad{\h}}\right)}{\left(1 + \left|\takegrad{\h}\right|^2\right)^2} \approx \det\left(\takegrad{\takegrad{\h}}\right),
    \end{align}
    \end{subequations}
    where derivatives are taken with respect to surface membrane parameters, $\left(\U, \V\right)$, and the approximation is justified because, physically, we expect $\left|\takegrad{\h}\right| \ll 1$.
\end{enumerate}
Then the partition function, $\partFunc$, is obtained by integrating over kinematically admissible deformations,
\begin{equation} \label{eq:part1}
    \partFunc = \int \exp\left(-\Ham\left[\hCons\right] / \kB \T\right) \Df{\hCons},
\end{equation}
where $\Ham\left[\hCons\right] = \Hcrease + \HvdW + \Hbend + \HGauss$ is the energy of a microstate and the tilde is used to indicate that the out of plane membrane deformations, $\hCons$, are generally subject to constraints.
For the present case, the constraints include \begin{inparaenum}[1)] \item the top and bottom membranes should not pass through each other--the ``self-contact constraint''--and \item boundary conditions along the edge where the top and bottom membranes meet the crease (i.e. $\U = 0$). \end{inparaenum}
With the partition function in hand, the probability of a microstate is given by
\begin{equation} \label{eq:prob}
    \prob\left[\h\right] = \frac{1}{\partFunc} \exp\left(-\Ham\left[\h\right] / \kB \T\right),
\end{equation}
and the Helmholtz free energy by
\begin{equation} \label{eq:free-energy}
    \freeEnergy = -\kB \T \log \partFunc.
\end{equation}

\begin{figure*}
	\centering
	\includegraphics[width=\linewidth]{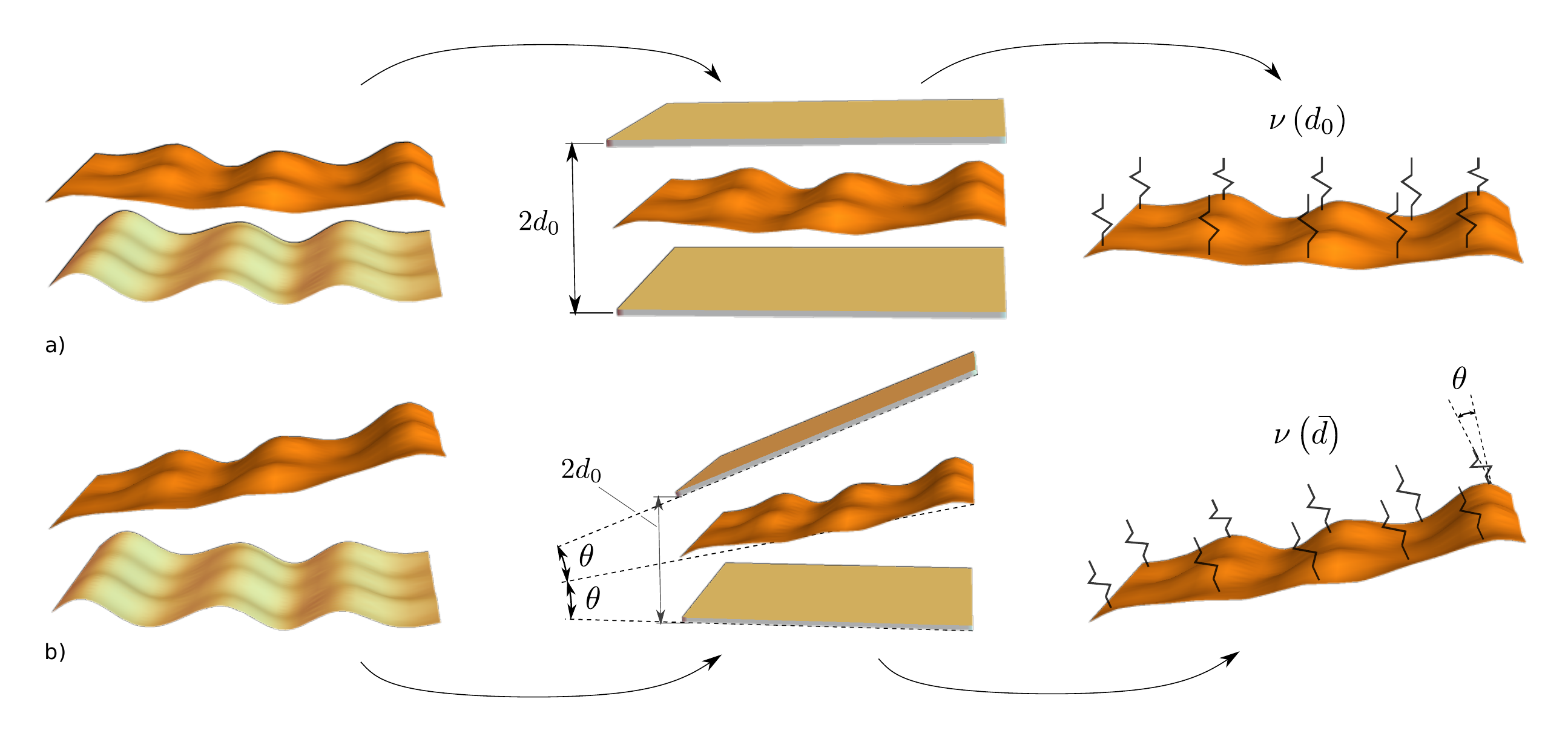}
	\caption{
		\capTitle{Modified Helfrich-like approach for two membranes at an angle with respect to each other.}
		a) Helfrich's approach to model the entropic pressure between two \emph{parallel} membranes separated by a mean distance $\DO$.
		This case is equivalent to a single fluctuating membrane between two hard walls that are separated by a distance $2\DO$.
		The hard wall constraints are weakly enforced by a harmonic potential--the strength of which depends on $\DO$.
		b) Extension of approach to membranes at an angle, $\creaseAngle$.
		Mapping to a single fluctuating membrane at two hard walls, each at angle $\creaseAngle$, to the membrane.
		The hard wall constraints are weakly enforced by a harmonic potential which depends on $\DAvg$ and where the force of the potential is projected orthogonal to the hard walls.
	}
	\label{fig:Helfrich-mapping}
\end{figure*}

In its present form, \eqref{eq:part1} is prohibitively difficult to evaluate analytically.
To make further progress, we take inspiration from seminal work by Helfrich~\cite{helfrich1978steric} where it was shown that the problem of two \emph{parallel} fluctuating membranes separated by a mean distance $\DO$ is equivalent to a single fluctuating membrane between two hard walls that are separated by a distance $2\DO$ (see \fref{fig:Helfrich-mapping}.a).
Helfrich then simplified further by replacing the hard walls with a harmonic potential on the out of plane undulations with spring constant $\vPot$ per unit area.
The magnitude of $\vPot$ was then chosen to (energetically) constrain the average undulations such that $\avg{\h^2} < \DO^2$; where by $\avg{.}$, we mean the thermodynamic average so that
\begin{equation} \label{eq:hsqrd}
    \avg{\h^2} = \int \h^2 \prob\left[\h\right] \Df{\h} = \frac{1}{\partFunc} \int \h^2 \exp\left(-\Ham\left[\h\right] / \kB \T\right) \Df{\h}.
\end{equation}
Here, as shown in \fref{fig:Helfrich-mapping}.b, we propose a similar mapping where two fluctuating membranes separated by a minimum distance $\DO$ and with dihedral angle $\creaseAngle$ is mapped to a single fluctuating membrane between two hard walls each with minimum distance $\DO$ from the membrane and with dihedral angles $\creaseAngle$.
The hard walls are then subsequently modeled via a harmonic potential $\vPot\left(\DAvg\right)$ that serves to energetically impose the constraint $\avg{\left(\h \cos \creaseAngle\right)^2} < \DAvg^2$ where the factor $\cos \creaseAngle$ appears because the out of plane displacement field $\h$ has a component of $\pm \h \cos \creaseAngle$ towards the top and bottom walls, respectively.
The top and bottom folded membranes become free to fluctuate as $\creaseAngle \rightarrow \pi / 2$, as expected.
\hl{Note, although the distance between the two membranes varies with $\U$, a mean-field approximation is taken and it is assumed that the confining potential, $\vPot$, is constant in space\footnote{
	\hl{
		Mean-field approximations are a common practice in statistical mechanics for achieving analytical tractability. Investigation
of this argument via high fidelity molecular dynamics, other molecular simulation techniques, or variational approximation
along the lines of~\cite{ahmadpoor2017thermal}, presents a potentially interesting topic for future research.
	}
}.}

\subsection{Freely fluctuating, parallel membranes revisited}
As an illustration, we first revisit the problem of freely fluctuating, parallel membranes separated by a distance $\D$.
For this case, the partition function is given by
\begin{equation} \label{eq:part-free}
	\partFunc_f = \int \exp\left(-\frac{\Hbend\left[\h\right] + \HGauss\left[\h\right] + \frac{1}{2} \vPot_f \intOverSurf{\h^2}}{\kB \T}\right) \Df{\h}
\end{equation}
where, for now, we have dropped the energy of the crease and the var der Waals potential, and where the subscript $f$ is used to signify that this is the ``free'' case.
To perform the functional integration over $\h$, it is convenient to choose a countable basis.
Here we choose to represent $\h$ with a Fourier series because it diagonalizes the Laplacian operator and, consequently, leads to independent Gaussian integrals over each of the Fourier coefficients.
Next let
\begin{equation}
	\qm = \frac{2 \pi m}{\Lu}, \quad \pn = \frac{2 \pi n}{\Lv},
\end{equation}
where
\begin{equation}
	\left(m, n\right) \in \modeNums = \left\{\Nats \times \Nats : \sqrt{\qm^2 + \pn^2} \leq \frac{2 \pi}{\lt}\right\},
\end{equation}
and $\Nats$ is the set of natural numbers.
Then
\begin{equation} \label{eq:h-Fourier}
\begin{split}
	\h = \sum_{\left(m, n\right) \in \modeNums} \Big\{ &\Amn \cqm \cpn + \Bmn \cqm \spn \\ & + \Cmn \sqm \spn + \Dmn \sqm \cpn \Big\},
 \end{split}
 \end{equation}
 and,
 \begin{subequations}
\begin{align}
	\frac{1}{2} \vPot_f\intOverSurf{\h^2} &= \sum_{\left(m, n\right) \in \modeNums} \left\{ \frac{\area \vPot_f}{8} \left(\Amn^2 + \Bmn^2 + \Cmn^2 + \Dmn^2\right) \right\}, \\
	\Hbend &= \sum_{\left(m, n\right) \in \modeNums} \left\{ \frac{\area \kbend}{8} \left(\qm^2 + \pn^2\right) \left(\Amn^2 + \Bmn^2 + \Cmn^2 + \Dmn^2\right)\right\}, \\
    \label{eq:Gauss-energy}
	\HGauss &= 0.
\end{align}
\end{subequations}
That the energy due to Gaussian curvature vanishes for periodic $\h$ is a straightforward consequence of the Gauss-Bonnet theorem.
Substituting into \eqref{eq:part-free},
\begin{equation}
\begin{split}
    \partFunc_f = \prod_{\left(m, n\right) \in \modeNums} \Bigg\{ &\int_{-\infty}^\infty \exp\left(-\frac{\area \left(\vPot_f + \kbend\left(\qm^2 + \pn^2\right)\right) \Amn^2}{8 \kB \T}\right) \df{\Amn} \\ &\dots \int_{-\infty}^\infty \exp\left(-\frac{\area \left(\vPot_f + \kbend\left(\qm^2 + \pn^2\right)\right) \Dmn^2}{8 \kB \T}\right) \df{\Dmn} \Bigg\}
\end{split}
\end{equation}
which is a product of independent Gaussian integrals over each of the Fourier coefficients. 
This is easily evaluated as
\begin{equation} \label{eq:part-free-eval}
    \partFunc_f = \prod_{\left(m, n\right) \in \modeNums} \left(\frac{8 \pi \kB \T}{\area \left(\vPot_f + \kbend\left(\qm^2 + \pn^2\right)\right)}\right)^2,
\end{equation}
and, using \eqref{eq:hsqrd}, we have the result
\begin{equation}
    \avg{\h^2}_f = \frac{\kB \T}{8 \sqrt{\vPot_f \kbend}}.
\end{equation}
Recall that the purpose of the harmonic potential $\int \vPot_f \h^2 / 2$ is to energetically constrain the membranes from contacting each other.
To this end, we define the factor $\stericFactor \in \Reals$ such that $\stericFactor \D^2 = \avg{\h^2}$.
Although $\stericFactor$ is unknown in general, enforcing $\stericFactor < 1$ ensures that $\avg{\h^2} < \D^2$, as desired.
Therefore, the harmonic potential constant is
\begin{equation}
    \vPot_f = \frac{1}{\kbend} \left(\frac{\kB \T}{8 \stericFactor \D^2}\right)^2.
\end{equation}

The free energy of the system can now be obtained from \eqref{eq:free-energy}
\begin{equation} \label{eq:free-energy-free}
    \begin{split}
        \freeEnergy_f &= -2 \kB \T \sum_{\left(m, n\right) \in \modeNums} \log \frac{8 \pi \kB \T}{\area \left(\vPot_f + \kbend\left(\qm^2 + \pn^2\right)\right)} \\
        &= 2 \kB \T \sum_{\left(m, n\right) \in \modeNums} \log \left(\frac{1}{\kbend} \left(\frac{\kB \T}{8 \stericFactor \D^2}\right)^2 + \kbend\left(\qm^2 + \pn^2\right)\right) + \freeEnergyConst \\
        &= \frac{\area}{64 \kbend \stericFactor} \left(\frac{\kB \T}{\D}\right)^2,
    \end{split}
\end{equation}
where, in the last step, the sum is approximated by an integral\footnote{The integral is approximated by converting the wave vector to polar coordinates and extending the bounds of integration such that the radial distance is integrated from $0$ to $\infty$ and the azimuth angle from $0$ to $2 \pi$.} and we have dropped the constant term, $\freeEnergyConst$, because only differences in free energy are physically meaningful.
The above implies an entropic pressure between the two membranes which scales as $1 / \D^3$; this, along with results of \eqref{eq:part-free-eval}--\eqref{eq:free-energy-free}, are well known~\cite{helfrich1978steric} and are consistent with experimental observations of the interactions between membranes (interactions which are otherwise difficult to explain).
Lastly, Helfrich~\cite{helfrich1978steric,helfrich1984undulations} was able to approximate $\stericFactor$ by averaging two limiting behaviors: \begin{inparaenum}[1)]
\item considering the constraint of $-\D < \h < \D$ on only a small number of points and \item constraining each material point of the membranes while also allowing one specific mode of deformation. \end{inparaenum}
The result is $\stericFactor = 1 / 6$.

\subsection{Constrained membranes at an angle}
Here we follow a similar approach while also considering the nuances associated with having the two membranes joined by a crease and oriented at some angle with respect to each other, $\creaseAngle$ (such as in \fref{fig:Helfrich-mapping}.b).
Let the $\left(\U, \V\right)$ parametrization be such that the top membrane meets the crease at $\U = 0$.
This implies $\h\left(0, \V\right) = 0$.
Also recall that the projection of the out of plane deformations of the top membrane towards the bottom membrane is $\h \cos \creaseAngle$.
The partition function is therefore
\begin{equation} \label{eq:part-fold}
	\partFunc = \exp\left(-\frac{\Hcrease + \HvdW}{\kB \T}\right) \int \exp\left(-\frac{\Hbend\left[\h\right] + \HGauss\left[\h\right] + \frac{1}{2} \vPot \intOverSurf{\left(\h \cos \creaseAngle\right)^2}}{\kB \T}\right) \diracFunctional{\h\left(0, \V\right)} \Df{\h},
\end{equation}
where $\diracFunctional{.}$ is the Dirac functional.
Following~\cite{fredrickson2006equilibrium,grasinger2021flexoelectricity} (and others), we represent the Dirac functional using its Fourier transform:
\begin{equation}
    \diracFunctional{\h\left(0, \V\right)} = \int \exp\left(-i \int_0^{\Lv} \df{\V} \: \waveFunc\left(\V\right) \h\left(0, \V\right) \right) \Df{\waveFunc}
\end{equation}
where $\waveFunc$ is a function which is the infinite dimensional analog of a wave vector.
Here we choose to represent $\waveFunc$ in the same basis as $\h$ such that
\begin{align}
    \waveFunc\left(\V\right) &= \sum_{k = 0}^\infty \Wk \cpn + \Vk \spn, \\
    \h\left(0, \V\right) &= \sum_{\left(m, n\right) \in \modeNums} \left\{ \Amn \cpn + \Bmn \spn\right\},
\end{align}
and, consequently,
\begin{equation}
\begin{split}
    \int_0^{\Lv} \df{\V} \: \waveFunc\left(\V\right) \h\left(0, \V\right) &= \frac{\Lv}{2} \sum_{\left(m, n\right) \in \modeNums} \left\{ \sum_{k = 0}^\infty  \left(\diracDelta{kn} \Amn \Vk + \diracDelta{kn} \Bmn \Wk \right)\right\} \\
    &= \frac{\Lv}{2} \sum_{\left(m, n\right) \in \modeNums} \left\{ \Amn \Vn + \Bmn \Wn \right\}.
\end{split}
\end{equation}
For brevity, let
\begin{equation}
	\Hmn = \frac{\area}{8} \left(\vPot \cos^2 \creaseAngle + \kbend\left(\qm^2 + \pn^2\right)\right).
\end{equation}
Then the partition function is
\begin{equation}
\begin{split}
    \partFunc = \exp\left(-\frac{\Hcrease + \HvdW}{\kB \T}\right) \prod_{\left(m, n\right) \in \modeNums} \Bigg\{ &\int_{-\infty}^\infty \int_{-\infty}^\infty \exp\left(-\frac{\Hmn \Amn^2}{\kB \T} - i \frac{\Lv}{2} \Vn \Amn \right) \df{\Amn} \df{\Vn} \\ & \times \int_{-\infty}^\infty \int_{-\infty}^\infty \exp\left(-\frac{\Hmn \Bmn^2}{\kB \T} - i \frac{\Lv}{2} \Wn \Bmn \right) \df{\Bmn} \df{\Wn} \\ &\times \int_{-\infty}^\infty \exp\left(-\frac{\Hmn \Cmn^2}{\kB \T}\right) \df{\Cmn} \int_{-\infty}^\infty \exp\left(-\frac{\Hmn \Dmn^2}{\kB \T}\right) \df{\Dmn} \Bigg\}
\end{split}
\end{equation}

As a result of the constraint, the integrals over $\Amn$ and $\Bmn$ take a different form but are still Gaussian.
Here
\begin{equation}
\begin{split}
    \int_{-\infty}^\infty \int_{-\infty}^\infty \exp\left(-\frac{\Hmn \Amn^2}{\kB \T} - i \frac{\Lv}{2} \Vn \Amn \right) \df{\Amn} \df{\Vn} &= \sqrt{\frac{\pi \kB \T}{\Hmn}} \int_{-\infty}^\infty \exp\left(-\left(\frac{\Lv}{2}\right)^2 \frac{\Vn^2}{\left(4 \Hmn / \kB \T\right)}\right) \df{\Vn} \\
    &= \sqrt{\frac{\pi \kB \T}{\Hmn}} \sqrt{\frac{4 \pi \Hmn}{\left(\Lv / 2\right)^2 \kB \T}} \\
    &= \frac{4 \pi}{\Lv},
\end{split}
\end{equation}
which, notably, is invariant with respect to $\Hmn$ and, as a consequence, $\creaseAngle$, $\vPot$, $\kbend$, and $\kB \T$.
This is likewise true for the integration with respect to $\Bmn$.
After taking the logarithm of $\partFunc$, these factors would simply become additive constants to the free energy, so we drop them here.
Thus, \emph{the displacement constraint at the crease effectively removes the $\cqm$ modes from the undulations of the top and bottom membranes}.
This is intuitive upon inspection of the Fourier series representation of $\h$, \eqref{eq:h-Fourier}, because, for the constraint $\h\left(0, \V\right)$ to be satisfied for all $\V$, is it is necessary that $\Amn = \Bmn = 0$ for all $\left(m, n\right) \in \modeNums$.

The partition function is finally evaluated as
\begin{equation} \label{eq:part-fold-eval}
    \partFunc = \exp\left(-\frac{\Hcrease + \HvdW}{\kB \T}\right) \prod_{\left(m, n\right) \in \modeNums} \left(\frac{8 \pi \kB \T}{\area \left(\vPot \cos^2 \creaseAngle + \kbend\left(\qm^2 + \pn^2\right)\right)}\right),
\end{equation}
where we note that, in comparing with \eqref{eq:part-free-eval}, \begin{inparaenum}[1)] \item each term in the product has half the exponent as the free membranes case because only half the modes contribute to the product, 
and \item the $\vPot$ factor in the denominator of each term now has the form $\vPot \cos^2 \creaseAngle$ because the membranes are at the angle $\creaseAngle$ with respect to each other. \end{inparaenum}
Using \eqref{eq:hsqrd}, we have the result
\begin{equation} \label{eq:hsqrdCons}
    \avg{\h^2 \cos^2 \creaseAngle} = \frac{\kB \T \cos \creaseAngle}{16 \sqrt{\vPot \kbend}}.
\end{equation}
Analogous to the free case, let $\avg{\h^2 \cos^2 \creaseAngle} = \stericFactor \DAvg^2$ and
\begin{equation}
    \vPot = \frac{1}{\kbend} \left(\frac{\kB \T \cos \creaseAngle}{16 \stericFactor \DAvg^2}\right)^2.
\end{equation}

The free energy of the system can now be obtained from \eqref{eq:free-energy}
\begin{equation} \label{eq:free-energy-fold0}
    \begin{split}
        \freeEnergy &= \Hcrease + \HvdW -\kB \T \sum_{\left(m, n\right) \in \modeNums} \log \frac{8 \pi \kB \T}{\area \left(\vPot \cos^2 \creaseAngle + \kbend\left(\qm^2 + \pn^2\right)\right)} \\
        &= \Hcrease + \HvdW + \kB \T \sum_{\left(m, n\right) \in \modeNums} \log \left(\vPot \cos^2 \creaseAngle + \kbend\left(\qm^2 + \pn^2\right)\right) + \freeEnergyConst \\
        &= \Hcrease + \HvdW + \underbrace{\frac{\area}{256 \kbend \stericFactor} \left(\frac{\kB \T \cos \creaseAngle}{\DAvg}\right)^2}_{\freeEnergyH},
    \end{split}
\end{equation}
where the contribution to the free energy due to undulations, $\h$, is denoted by $\freeEnergyH$ and, again, the constant term $\freeEnergyConst$ is dropped in the last step.
Given \eqref{eq:free-energy-fold0}, some remarks are in order.
Recall that for the freely fluctuating, parallel membrane case, the undulations of the two membranes lead to an entropic pressure.
Similarly, we see from \eqref{eq:free-energy-fold} that the undulations cause an entropic torque about the crease:
\begin{equation} \label{eq:entropic-torque}
    \torqueH = -\frac{\partial \freeEnergyH}{\partial \creaseAngle} = \frac{\area \left(\kB \T\right)^2 \left(\frac{\Lu}{2} + \DAvg \sin \creaseAngle \cos \creaseAngle \right)}{128 \kbend \stericFactor \DAvg^3}.
\end{equation}
As expected, we see that the entropic torque vanishes as $\creaseAngle \rightarrow \pi / 2$.
This is because the undulations of the two membranes do not interact with each other when the crease opens up.
Similar to the entropic pressure of the freely fluctuating, parallel membranes, the entropic torque of the folded membranes also scales as $\left(\kB \T\right)^2 / \DO^3 \kbend \stericFactor$.
\hl{The first term in the parentheses of the numerator includes a factor of $\Lu / 2$, which can be understood as the entropic torque being like an entropic pressure (from half the undulation modes of the free case) torquing the crease with a lever arm of $\Lu / 2$.
The second term (in the parentheses) is due to the dependence of $\DAvg$ on $\creaseAngle$.}
An equivalent result can be obtained by a different approach where the constraint at the crease is instead enforced via an energy penalty.
For completeness, this approach is outlined in \fref{app:constraint2}.

Finally, before proceeding, we extend and refine \eqref{eq:free-energy-fold0} (which was derived by assuming that $-\pi / 2 \leq \creaseAngle \leq \pi / 2$) by recognizing that \begin{inparaenum} [1)] \item the entropic repulsion and van der Waals attraction should both vanish when $\creaseAngle > \pi / 2$, and \item hard contact will occur between the top and bottom membrane for some $\creaseAngle \leq 0$. \end{inparaenum}
Therefore, we use the form
\begin{equation} \label{eq:free-energy-fold}
    \freeEnergy = \begin{cases}
        \Hcrease + \HvdW + \freeEnergyH,  &\creaseAngleClosed \leq \creaseAngle \leq \frac{\pi}{2} \\
        \Hcrease, &\: \frac{\pi}{2} < \creaseAngle \leq \pi.
    \end{cases}
\end{equation}
which is both continuous and consistent with the aforementioned expectations.
We determine $\creaseAngleClosed$ by solving for $\creaseAngle$ when the edge of the top membrane, $\U = \Lu$, comes into contact, on average, with the bottom membrane:
\begin{equation}
\begin{split}
	0 &= \DO - \Lu \tan \creaseAngleClosed - \sqrt{\avg{\left(\h \cos \creaseAngleClosed\right)^2}}\\
	  &= \DO - \Lu \tan \creaseAngleClosed - \sqrt{\stericFactor} \left(\DO + \frac{\Lu}{2} \tan \creaseAngleClosed\right),
\end{split}
\end{equation}
which results in
\begin{equation} \label{eq:creaseAngleClosed}
	\creaseAngleClosed = -\arctan\left(\frac{2 \DO \left(1 - \sqrt{\stericFactor}\right)}{\Lu\left(2 + \sqrt{\stericFactor}\right)}\right).
\end{equation}
In the limit of vanishing membrane undulations, $\stericFactor \rightarrow 0$, this recovers the angle at which zero temperature contact occurs: $\creaseAngleClosed \rightarrow -\arctan\left(\DO / \Lu\right)$.
Because the form of the free energy is different in each, we refer to $\left[\creaseAngleClosed, \pi / 2\right]$ as the \emph{folded interval}, to $\creaseAngle > \pi / 2$ as being ``unfolded'', and refer to $\creaseAngle = \pi$ as ``flat''.

\section{Folded stability}

Equilibrium states of the system correspond with local extrema of the free energy.
A state is stable if the free energy is locally increasing, and its unstable if there are local perturbations of the system for which the free energy is decreasing.
The primary kinematic description of the folded membrane is $\creaseAngle$; thus, $\partialInline{\freeEnergy}{\creaseAngle} = 0$ is a sufficient condition for a state to be in equilibrium. 
To this end, we consider roots of
\begin{equation} \label{eq:equilibrium}
	\partialx{\freeEnergy}{\creaseAngle} = \begin{cases} \kcrease \Lv \left(\creaseAngle - \creaseAngleZero\right) + \frac{\CvdW \area \Lu \sec^2 \creaseAngle}{12 \pi} \left(\frac{1}{\DAvg^3} - \frac{2}{\left(\lt + \DAvg\right)^3} + \frac{1}{\left(2 \lt + \DAvg\right)^3}\right) 
	- \frac{\area \left(\kB \T\right)^2 \left(\frac{\Lu}{2} + \DAvg \sin \creaseAngle \cos \creaseAngle \right)}{128 \kbend \stericFactor \DAvg^3} &\creaseAngleClosed \leq \creaseAngle \leq \frac{\pi}{2}, \\
	\kcrease \Lv \left(\creaseAngle - \creaseAngleZero\right) & \: \frac{\pi}{2} < \creaseAngle \leq \pi. 
\end{cases}
\end{equation}
As in \eqref{eq:entropic-torque}, each term in the above equation can be seen as (the negative of) a thermodynamic torque.
The system is in equilibrium when the net torque vanishes.
We can make some general statements about each contribution to the torque.
Using \eqref{eq:entropic-torque}, we can derive the change in entropic torque as a function of $\creaseAngle$,
\begin{equation}
	\partialx{\torqueH}{\creaseAngle} = \frac{\area \left(\kB \T\right)^2 \left(4 \DAvg^2 \cos 2 \creaseAngle - \Lu \left(3 \Lu \sec^2 \creaseAngle + 4 \DAvg \tan \creaseAngle\right)\right)}{512 \kbend \stericFactor \DAvg^4}.
\end{equation}
The sign of this term is only a function of $\Lu$, $\DO$, and $\creaseAngle$.
When $\DO \leq \Lu \sqrt{3} / 2$, this is nonpositive in the folded interval.
Thus, the entropic torque is nonnegative, monotonically decreasing, and approaches $0$ as $\creaseAngle \nearrow \pi / 2$.
The torque due to the van der Waals potential is nonpositive and also approaches $0$ as $\creaseAngle \nearrow \pi / 2$.
Finally, the torque due to the crease energy is linear in $\creaseAngle$; clearly, it is monotonically decreasing and it is positive (negative) when $\creaseAngle < \creaseAngleZero$ (when $\creaseAngle > \creaseAngleZero$).
Because of what is known about the signs of the various contributions to the thermodynamic torque, we can conclude that:
\begin{enumerate}[1)]
	\item when $\CvdW = 0$, equilibrium states can only exist when $\creaseAngle \geq \creaseAngleZero$.
	Further, the system has a single stable, equilibrium state when $\creaseAngleZero \geq \pi / 2$.
	\item similarly, when $\left(\kB \T\right)^2 / \kbend = 0$, equilibrium states can only exist when $\creaseAngle \leq \creaseAngleZero$.
\end{enumerate}

\paragraph*{Aspect ratio.}
Next we consider the role of the dimensions of the folded sheet on its stability.
Let $\aspectRatio = \Lv / \Lu$.
Then
\begin{equation}
    \lim_{\aspectRatio \rightarrow \infty} \freeEnergy = \frac{1}{2} \kcrease \Lv \left(\creaseAngle - \creaseAngleZero\right)^2
\end{equation}
such that the origami system is monostable and its only minima is at $\creaseAngle = \creaseAngleZero$.
An equivalent result is obtained when $\kcrease \rightarrow \infty$.

In contrast, consider $\aspectRatio \rightarrow 0$.
For this case, when $\creaseAngle < \pi / 2$, the crease energy is negligible compared to the van der Waals and entropic contributions.
Then, an equilibrium state in the folded interval exists provided there is a state where the entropic torque balances the torque due to the van der Waals attraction.
While it is difficult to say with generality what the stability properties of the system are in the folded interval, given the continuity of the torques and that they vanish as $\creaseAngle \nearrow \pi / 2$, we have that $\creaseAngle = \pi / 2$ is an equilibrium state.
It is stable (unstable) from below provided that there are an odd (even) number of equilibria on the folded interval.
Consider now two separate cases for the crease energy.
If $\creaseAngle < \pi / 2$, then $\creaseAngle = \pi / 2$ stable from above and so the system is multistable provided there are two or more equilibrium states in the folded interval. If $\creaseAngleZero > \pi / 2$ and $\kcrease > 0$, then $\creaseAngle = \pi / 2$ is unstable from above (and, by definition, \emph{unstable}) and $\creaseAngle = \creaseAngleZero$ is stable.
Then the system is multistable provided there are one or more equilibrium states in the folded interval.
The number of equilibrium states in the folded interval will depend nontrivially on the competition between entropic torque, which scales as $\left(\kB \T\right)^2 / \kbend$, and the torque due to van der Waals attraction.

\paragraph*{Separation of spatial scales.}
To make further progress, we consider the scaling: $\lt \ll \DO \ll \min\left\{\Lu, \Lv\right\}$.
First, we expand $\partial \freeEnergy / \partial \creaseAngle$ in powers of $\lt / \DAvg$:
\begin{equation}
	\partialx{\freeEnergy}{\creaseAngle} = \Lv \kcrease \left(\creaseAngle - \creaseAngleZero\right) -\frac{\area \left(\kB \T\right)^2 \sin 2 \creaseAngle}{256 \: \DAvg^2 \kbend \stericFactor} + \frac{\Lu^2 \Lv}{\DAvg^3} \left(\frac{\CvdW \left(\frac{\lt}{\DAvg}\right)^2 \sec^2 \creaseAngle}{\pi} - \frac{\left(\kB \T\right)^2}{256 \: \kbend \stericFactor}\right) + \orderOf{\left(\frac{\lt}{\DAvg}\right)^3}.
\end{equation}
Next, assume that $\creaseAngle$ is small enough such that terms $\orderOf{\left(\min\left\{\Lu, \Lv\right\} / \DAvg\right)^3}$ dominate.
Then, to a good approximation,
\begin{equation} \label{eq:equilibriumApprox}
	\partialx{\freeEnergy}{\creaseAngle} \approx \frac{\Lu^2 \Lv}{\DAvg^3} \left(\frac{\CvdW \left(\frac{\lt}{\DAvg}\right)^2 \sec^2 \creaseAngle}{\pi} - \frac{\left(\kB \T\right)^2}{256 \: \kbend \stericFactor}\right),
\end{equation}
	and,
\begin{equation} \label{eq:thetaStar1}
	\creaseAngle = \pm \arccos \left(\frac{16 \left(\lt / \DAvg\right) \sqrt{\CvdW}}{\kB \T \sqrt{\pi / \kbend \stericFactor} \DAvg}\right) \implies \partialx{\freeEnergy}{\creaseAngle} \approx 0.
\end{equation}
The above equation, however, is still implicit and nonlinear.
Expanding $\DAvg$ to linear order in $\creaseAngle$, we finally obtain:
\begin{equation} \label{eq:equilibria}
	\creaseAnglePM = \frac{\pm \arccos \vdWToEnt}{1 - \frac{\Lu \vdWToEnt}{2 \DO \sqrt{1 - \vdWToEnt^2}}}
\end{equation}
as approximate equilibrium states, $\creaseAngleL$ and $\creaseAngleP$, where
\begin{equation}
	\vdWToEnt = \frac{\left(\lt / \DO\right) \sqrt{\CvdW / \pi}}{\kB \T / \left(16 \sqrt{\kbend \stericFactor}\right)}
\end{equation}
is a measure of the ratio of the competition between the van der Waals attraction between the two membranes and entropic repulsion.
Based on the arguments to $\arccos$ and the square root, for these approximations to equilibrium states to be real, 
we require that
\begin{equation} \label{eq:thetaStar1Cond1}
	\frac{\kB \T}{16 \sqrt{\kbend \stericFactor}} \geq \frac{\lt}{\DO} \sqrt{\frac{\CvdW}{\pi}};
\end{equation}
that is, that the thermal undulations are not overcome by the van der Waals attraction.
If \eqref{eq:thetaStar1Cond1} holds, then, upon inspection of \eqref{eq:equilibriumApprox}, we see that
\begin{equation}
	\begin{cases}
		\partialx{\freeEnergy}{\creaseAngle} \lessapprox 0, & \text{if } \left|\creaseAngle\right| \lesssim \creaseAngleP \\
		\partialx{\freeEnergy}{\creaseAngle} \gtrapprox 0, & \text{if } \left|\creaseAngle\right| \gtrsim \creaseAngleP
	\end{cases}
\end{equation}
and, consequently, $\creaseAnglePM$ are local minima.
The symmetry of the above analysis about $\creaseAngle = 0$ is a direct consequence of the asymptotic limit of $\DO \ll \Lu$, because here the states $\creaseAngle$ and $-\creaseAngle$ are simply mirror images of each other.
As $\vdWToEnt \rightarrow 1$, a consequence is that $\creaseAnglePM \rightarrow 0$, which is reminiscent of a pitchfork bifurcation.
Another important consequence of this asymptotic limit is that $\creaseAngleClosed \sim 0$ and only $\creaseAngleP$ is accessible.
This leads to a system which is at least bistable with local minima at $\creaseAngle = \creaseAngleP$ and $\creaseAngleZero$, and local maxima at $\creaseAngle = 0$ and in the interval $\creaseAngle \in \left(\creaseAngleP, \creaseAngleZero\right)$.

\paragraph*{Strong van der Waals attraction.}
Here we investigate the possibility of equilibria near to, but also less than, $\creaseAngle = \pi / 2$. 
An approximation to $\partialInline{\freeEnergy}{\creaseAngle} |_{\creaseAngle} = 0$ can be obtained by Taylor expanding about $\creaseAngle = \pi / 2$ up to $\orderOf{\left(\pi / 2 - \creaseAngle\right)^3}$, and then solving the resulting cubic equation:
\begin{equation} \label{eq:cubic}
	-\Lv \kcrease \left(\creaseAngleZero - \frac{\pi}{2}\right) + \Lv \kcrease \left(\frac{\pi}{2} - \creaseAngle\right) + \frac{1}{16} \frac{\Lv}{\Lu} \left(\frac{512}{\pi} \left(\frac{\lt}{\Lu}\right)^2 \CvdW - \frac{\left(\kB\T\right)^2}{\kbend \stericFactor}\right) \left(\frac{\pi}{2} - \creaseAngle\right)^3 = 0.
\end{equation}
Let
\begin{align}
	p &= -\frac{16 \Lu \kcrease}{\frac{512}{\pi} \left(\frac{\lt}{\Lu}\right)^2 \CvdW - \frac{ \left(\kB\T\right)^2}{\kbend \stericFactor}}, \\
	q &= -\frac{16 \Lu \kcrease \left(\creaseAngleZero - \frac{\pi}{2}\right)}{\frac{512}{\pi} \left(\frac{\lt}{\Lu}\right)^2 \CvdW - \frac{\left(\kB\T\right)^2}{\kbend \stericFactor}}, \\
	\Delta &= \frac{q^2}{4} - \frac{p^3}{27},
\end{align}
then, provided $\Delta \geq 0$, the real root of \eqref{eq:cubic} is
\begin{equation} \label{eq:root}
	\frac{\pi}{2} - \creaseAngleStar = \left(\sqrt{\Delta} - \frac{q}{2}\right)^{1/3} - \left(\sqrt{\Delta} + \frac{q}{2}\right)^{1/3},
\end{equation}
where $\creaseAngleStar$ is the equilibrium state corresponding to the root.
Given \eqref{eq:cubic}-\eqref{eq:root}, we note the following:
\begin{enumerate}[1)]
	\item A necessary condition for the equilibrium state to be in the folded interval (i.e. $\creaseAngleStar \in \left[\creaseAngleClosed, \pi / 2\right]$) is that $q \leq 0$.
	Assuming $\creaseAngleZero \geq \pi / 2$, this implies
	\begin{equation} \label{eq:theta2Cond1}
		\frac{512}{\pi} \left(\frac{\lt}{\Lu}\right)^2 \CvdW \geq \frac{\left(\kB\T\right)^2}{\kbend \stericFactor}.
	\end{equation}
	Here, in contrast to the condition for equilibria near $\creaseAngle = 0$ (i.e. \eqref{eq:thetaStar1Cond1}), the van der Waals attraction, by some measure, must be sufficiently larger than the entropic repulsion.
	If instead $\creaseAngleStar > \pi / 2$, this would be inconsistent with the assumptions of \eqref{eq:cubic} because the van der Waals and entropic terms would necessarily vanish.
	\item The condition given by \eqref{eq:theta2Cond1} is also sufficient for the root being real.
	\item Any of the following limits:
	\begin{equation}
		\Lu \kcrease \rightarrow 0, \quad
		\left(\frac{\lt}{\Lu}\right)^2 \CvdW \rightarrow \infty, \quad
		\creaseAngleZero \rightarrow \frac{\pi}{2},
	\end{equation}
	implies that
	\begin{equation}
		\creaseAngleStar \rightarrow \frac{\pi}{2}.
	\end{equation}
	\item The nature of the van der Waals and entropic terms suggests the following: when $\left(\frac{\lt}{\Lu}\right)^2 \CvdW \rightarrow \infty$, $\creaseAngle \rightarrow \creaseAngleClosed$ is a minima and $\creaseAngleStar \rightarrow \pi / 2$ is a maxima.
\end{enumerate}

\paragraph*{Summary and hot limit.}
For either the case of \begin{inparaenum}[1)] \item $\lt \ll \DO \ll \min\left\{\Lu, \Lv\right\}$ and $\vdWToEnt \lesssim 1$, or \item $\left(\lt / \Lu\right)^2 \CvdW \rightarrow \infty$, the system is thermodynamically multistable provided that $\creaseAngleZero > \pi / 2$. \end{inparaenum}
However, it is clear (by \eqref{eq:equilibrium}) the system will be monostable in the high temperature limit, i.e. $\left(\kB \T\right)^2 / \kbend \stericFactor \rightarrow \infty$.
Here the stable state is $\creaseAngle = \creaseAngleZero$ when $\creaseAngleZero > \pi / 2$ and saturates at $\creaseAngle \rightarrow \pi / 2$ otherwise.

\subsection{Graphene with semi-cylindrical creases formed by elastic bending} \label{sec:grapheneEq}

A particular form of the crease torsional stiffness is of interest because of its simplicity.
Consider forming the crease by bending via a uniform, semi-cylindrical curvature for $\pi$ radians with a final radius of $\DO / 2$.
If done elastically, in terms of the bending stiffness, $\kbend$, we have that\footnote{
	\hl{The relationship between $\kcrease$ and $\kbend$ is determined by integrating the energy of mean curvature over the area of the crease, and equating it with \eqref{eq:creaseEnergy}.}
}
\begin{equation} \label{eq:elasticCreaseStiffness}
	\kcrease = \frac{2 \kbend}{\pi \DO}, \text{ and } \creaseAngleZero = \pi.
\end{equation}
Now we consider how the free energy landscape changes as a function of the geometry, mechanical properties, and temperature of the molecular origami.

\begin{figure*}
	\centering
	\includegraphics[width=\linewidth]{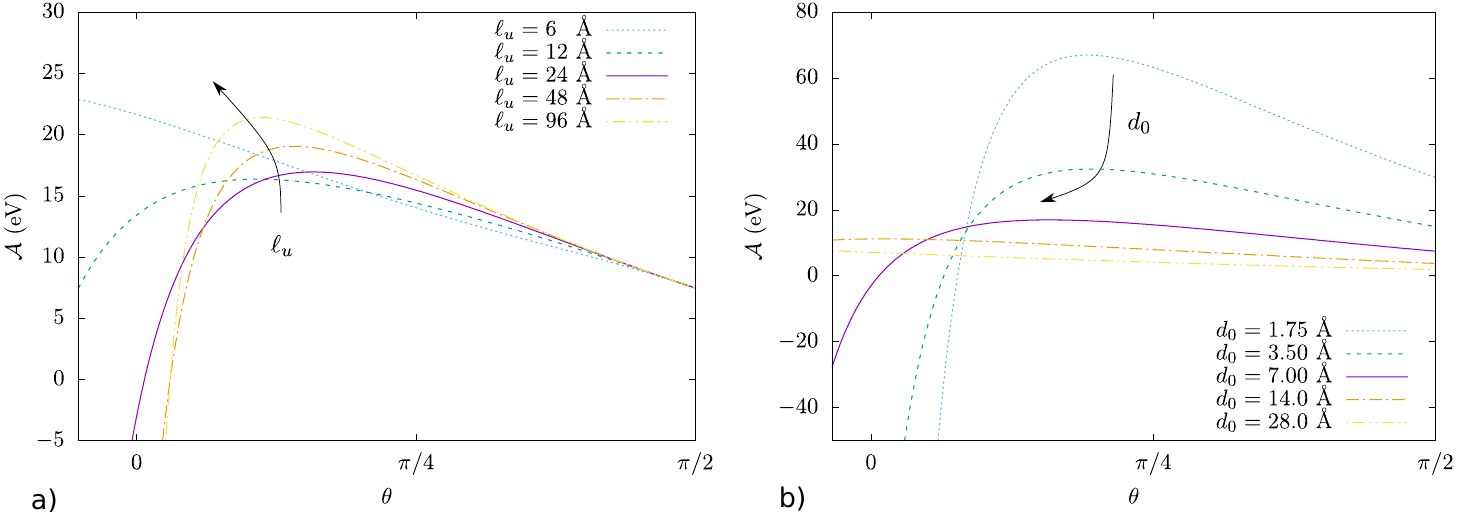}
	\caption{
		\capTitle{Free energy landscapes for varying geometry.}
		a) Free energy as a function of $\creaseAngle$ for varying $\Lu$.
		As $\Lu$ increases, the depth of the folded minima and the magnitude of the free energy barrier increase.
		When $\Lu$ drops below a critical length, the folded minima is annihilated via a saddle node bifurcation.
		b) Free energy as a function of $\creaseAngle$ for varying $\DO$.
		As $\DO$ increases, the depth of the folded minima and the magnitude of the free energy barrier decrease.
		When $\DO$ approaches a critical distance, the folded minima is annihilated via a saddle node bifurcation.
	}
	\label{fig:energy-landscapes_Lu-and-d0}
\end{figure*}

\begin{figure*}
	\centering
	\includegraphics[width=\linewidth]{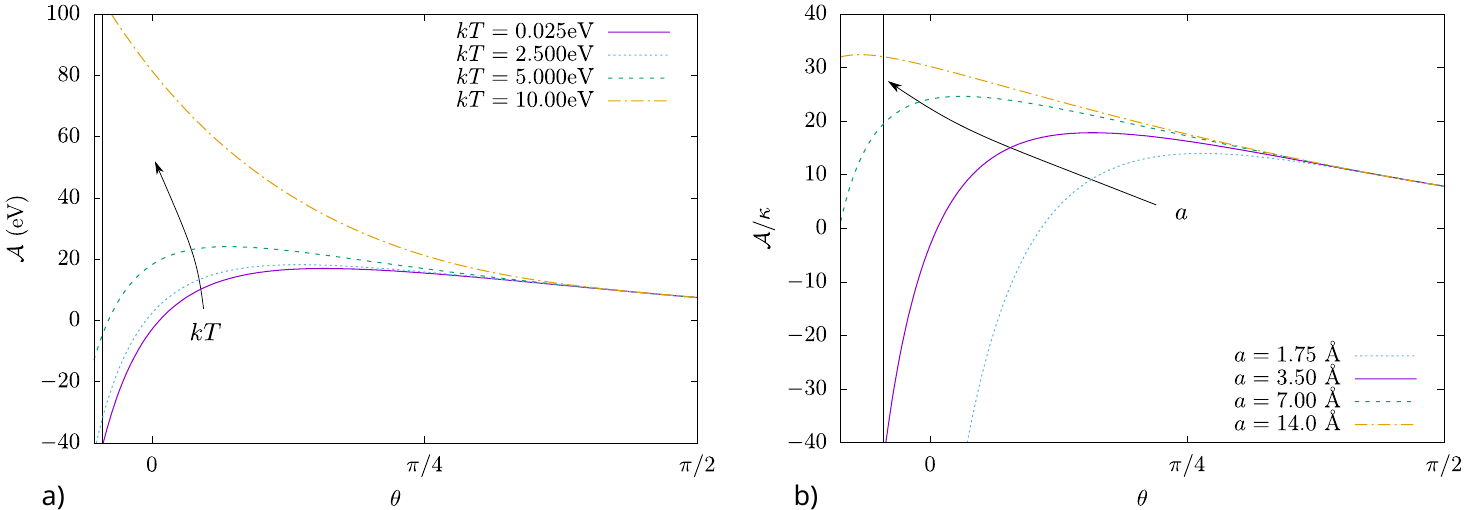}
	\caption{
		\capTitle{Free energy landscapes for varying temperature and membrane thickness.}
		a) Free energy as a function of $\creaseAngle$ for varying $\kB \T$.
		Increasing temperature leads to increasing entropic repulsion between the two membranes until the folded minima vanishes.
		b) Free energy as a function of $\creaseAngle$ for varying $\lt$; that is, varying number of stacked layers.
		Increasing $\lt$ simultaneously increases van der Waals attraction between the membranes and bending stiffness.
		The competition between the van der Waals attraction and crease energy moves the energy barrier up and to the left until the folded minima vanishes.
	}
	\label{fig:energy-landscapes_kT-and-thickness}
\end{figure*}

\begin{figure*}
	\centering
	\includegraphics[width=\linewidth]{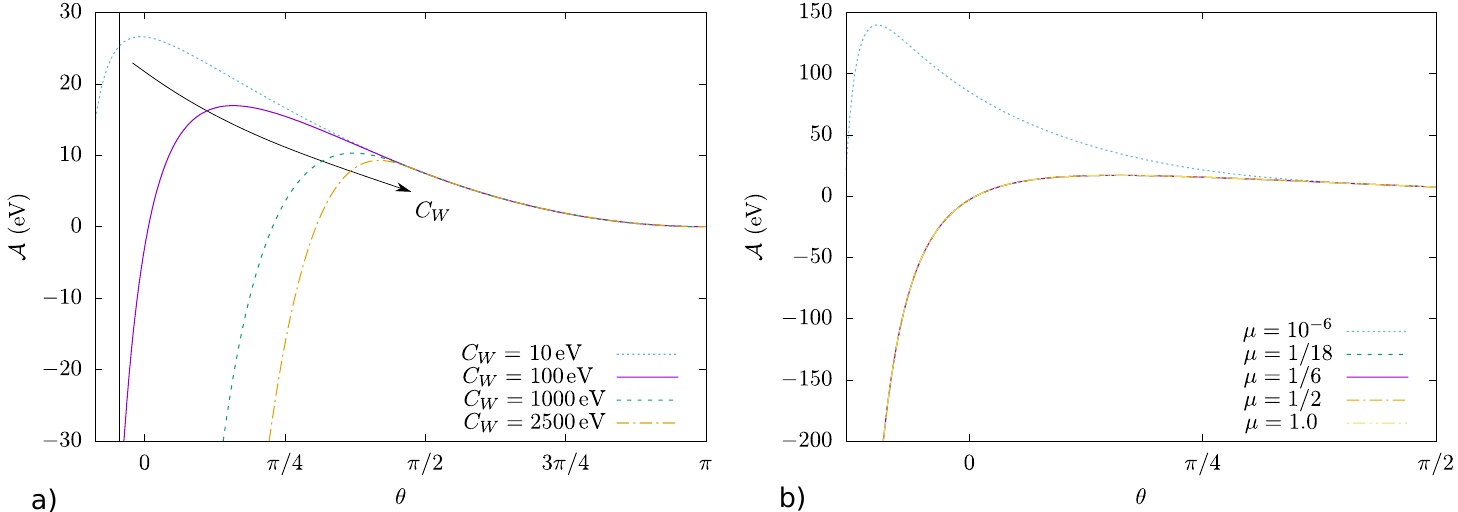}
	\caption{
		\capTitle{Free energy landscapes as a function of van der Waals coefficient and steric repulsion factor.}
		a) Free energy as a function of $\creaseAngle$ for varying $\CvdW$.
		As $\CvdW$ diverges, the energy barrier saturates at $\creaseAngle = \pi / 2$.
		b) Free energy as a function of $\creaseAngle$ for varying $\stericFactor$.
		For the system of interest, the free energy is insensitive to $\stericFactor$.
	}
	\label{fig:energy-landscapes_W-and-mu}
\end{figure*}

As a model system, we consider graphene with mechanical properties $\kbend = 0.95$ \eVSp and $\CvdW = 100$ \eV~\cite{yang2021energetics}.
The thickness of the graphene is assumed to be $\lt = 3.5$ \angs, and, following~\cite{yang2021energetics}, it is assumed that the crease has diameter, $\DO = 7$ \angs.
At zero temperature, the smallest square where the folded state has less energy than the flat state has a side length of approximately $70$ \angs~\cite{yang2021energetics}.
Here we probe the multistability of the system near this limiting case; therefore let $\Lv = 70$ \angs and $\Lu = 24$ \angs\footnote{The square is folded in half and some of the length is in the crease itself, $\Lv / 2 - \pi \DO / 2 \approx 24$ \angs}.
Following~\cite{mozaffari2021flexoelectricity}, let $\stericFactor = 1/6$.
At room temperature, $\kB \T = 0.025$ \eV.
In order to isolate the effect of each parameter, we vary it about this base system.
The results are presented in \fref{fig:energy-landscapes_Lu-and-d0}-\ref{fig:energy-landscapes_W-and-mu} where, for each figure, the base case is depicted by a solid purple line.

First, we consider the effects of $\Lu$ and $\DO$.
In \fref{fig:energy-landscapes_Lu-and-d0}.a, $\Lu$ is varied as $6$ \angs, $12$ \angs, $24$ \angs, $48$ \angs, and $96$ \angs, and the free energy as a function of $\creaseAngle$ is shown in the folded interval, $\left[\creaseAngleClosed, \pi / 2\right]$.
The free energy curves are identical for $\pi / 2 < \creaseAngle \leq \pi$ because the crease properties are the same; recall, there is a local minima at the flat state, $\creaseAngle = \pi$.
As $\Lu$ is increased, the free energy barrier between the folded minima and the flat state increases in magnitude and moves towards $\creaseAngle = 0$.
Note that \begin{inparaenum}[1)] \item the case of $\Lu = 24$ \angs, at finite temperature, though still stable, is at a higher free energy than the flat state, \item the difference in free energy between the minima increases for $\Lu = 12$ \angs, and \item the case of $\Lu = 6$ \angsSp is monostable; that is, the folded minima no longer exists as it has been annihilated via a saddle node bifurcation. \end{inparaenum}

Next consider folding the crease with varying radii of curvature.
In \fref{fig:energy-landscapes_Lu-and-d0}.b, $\DO$ is varied as $1.75$ \angs, $3.5$ \angs, $7$ \angs, $14$ \angs, and $28$ \angs.
As $\DO$ changes, there are various competing effects on the free energy landscape: the strength of the van der Waals attraction (for a given $\creaseAngle$) decreases, as does the entropic repulsion and effective crease stiffness, $\kcrease$ (via equation \eqref{eq:elasticCreaseStiffness}).
The combined effect is that the depth of the well for the folded minima decreases with increasing $\DO$, and the free energy barrier moves towards $\creaseAngle = 0$.
For increasing $\DO$, eventually the folded minima is annihilated via a saddle node bifurcation (similar to decreasing $\Lu$).

Next, in \fref{fig:energy-landscapes_kT-and-thickness}.a, $\kB \T$ is varied as $0.025$ \eV, $2.5$ \eV, $5$ \eV, and $10$ \eV.
As the temperature increases, the crease angles at which the minima and the energy barrier occur decrease until the minima is no longer accessible (at $\kB \T = 10$) due to self-contact, and the magnitude of the energy barrier increases.
The boundary of the accessible region is delimited by a solid, vertical line.
Eventually, thermal fluctuations overcome the van der Waals attraction.
This, however, only occurs at extreme temperatures because $\kbend$ and $\CvdW$ are both several orders of magnitude greater than $\kB \T$ at room temperature.

Multiple layers of a $2$D material may be stacked to increase its thickness.
Assume that if the thickness is rescaled by $\lt \rightarrow \zeta \lt$ then the bending stiffness is rescaled as $\kbend \rightarrow \zeta^3 \kbend$.
Stacking layers serves to increase the van der Waals attraction, and the increased bending stiffness makes the crease more stiff while simultaneously reducing the entropic repulsion.
In \fref{fig:energy-landscapes_kT-and-thickness}.b, the thickness, $\lt$, is varied from $1.75$ \angs, $3.5$ \angs, $7$ \angs, and $14$ \angs.
As the thickness increases, the energy barrier moves up and to the left until the system is monostable.
Here, in contrast to increasing temperature, the folded minima vanishes, not because of entropic repulsion, but because the crease energy dominates the van der Waals attraction.

Although the degree of tunability may be limited, we imagine altering the van der Waals attraction by functionalizing the surface of the graphene, or by other chemical alterations.
Here, in \fref{fig:energy-landscapes_W-and-mu}.a, the van der Waals coefficient, $\CvdW$ is varied from $10^1$ \eV, $10^2$ \eV, $10^3$ \eV, and $2.5 \times 10^3$ \eV.
The results here agree well with the previous analysis which suggested that making the van der Waals attraction sufficiently large causes the energy barrier to converge towards $\creaseAngle \rightarrow \pi / 2$ in the limiting case.
In contrast, as $\CvdW$ is decreased, the energy barrier increases in magnitude and occurs at lesser $\creaseAngle$ until the system is monostable (e.g. $\CvdW \lessapprox 10^1$ \eV).

Lastly, in \fref{fig:energy-landscapes_W-and-mu}.b, we probe the sensitivity of the steric repulsion factor, $\stericFactor$.
We see that, except for the case of vanishingly small $\stericFactor$, the energy landscapes are nearly identical.
Thus, at least for the model system considered herein, it is easy to justify $\stericFactor = 1/6$ as a reasonable choice.

\section{Thermally driven unfolding}

In the previous section, we investigated the \emph{thermodynamic equilibrium} multistability properties of a molecular origami system consisting of a single crease.
Although a folded stable state existed for many of the systems, the depth of the well varied and, in fact, was at a greater free energy than the flat state for the base system at any finite temperature.
Thus, although a folded state was stable, the flat state was energetically preferred; and the free energy barrier between them also varied, with a nontrivial dependence on the properties of the system.
Here we ask a different question: ``what is the long-time, nonequilibrium behavior of the system?''.
More specifically, ``when does the system prefer the flat state to any folded state?'', and ``assuming the system is initially in the folded state, at what rate will it unfold?''.
The latter question may be difficult to answer with a high degree of accuracy; however, tools originally developed for studying chemical reaction rates and derived from nonequilibrium statistical mechanics such as transition state theory (TST) and Kramers' escape rate theory~\cite{hanggi1990reaction,vanden2005transition,balakrishnan2008elements,bao2017variational}, will be used here to provide estimates.

To study the long-time, nonequilibrium behavior of the molecular origami system, we envision the following: the origami is surrounded by a fluid at a fixed temperature.
Thermal fluctuations of the surrounding fluid lead to collisions with the origami and act as a kind of stochastic torque on the crease, $\torqueSto\left(\tv\right)$.
The stochastic torque will generally cause changes to the crease angle, $\creaseAngle\left(\tv\right)$; however, we assume the time scale at which $\creaseAngle$ varies is much larger than the time scale of undulations of the top and bottom membranes.
Thus, we assume a kind of quasiequilibrium process where, for a given change in crease angle, the undulations of the membranes thermalize (effectively) instantaneously, and the free energy given by \eqref{eq:free-energy-fold} is a proper description of the state of the origami system\footnote{Recall that the free energy here is specific to the constant fold angle ensemble. We emphasize that this free energy does not correspond to a crease ergodically fluctuating through all allowable $\creaseAngle$.}.
The dynamics of the system can be described by the Langevin equation as
\begin{equation} \label{eq:langevin}
	\Imom \ddot{\creaseAngle} - \Imom \drag \dot{\creaseAngle} = -\partialx{\freeEnergy}{\creaseAngle} + \torqueSto,
\end{equation}
where $\dot{\generic} = \partialInline{\generic}{\tv}$, $\Imom$ is the moment of inertia of the top membrane (about the center of the crease), $\drag$ is the drag coefficient, and $-\partialInline{\freeEnergy}{\creaseAngle}$ acts as a thermodynamic torque.

Clearly, the Langevin equation given in \eqref{eq:langevin} is nonlinear and does not readily admit a solution.
However, it provides a formalism for estimating the rate of transition over an energy barrier for the case of when $\drag \tv \gg 1$, i.e. the diffusion regime.
Towards consideration of this regime, we drop the inertial term in \eqref{eq:langevin} and arrive at the Smoluchowski equation~\cite{balakrishnan2008elements,leadbetter2023statistical,hanggi1990reaction}.
Let $\prob\left(\creaseAngle, \tv\right)$ denote the probability density of a crease angle $\creaseAngle$ at time $\tv$.
The corresponding Fokker-Planck equation (of the Smoluchowski equation) is
\begin{equation} \label{eq:FPE}
	\partialx{\prob}{\tv} = \frac{1}{\Imom \drag} \partialx{ }{\creaseAngle}\left(\partialx{\freeEnergy}{\creaseAngle} \prob\right) + \frac{\kB \T}{\Imom \drag} \partialN{\prob}{\creaseAngle}{2}.
\end{equation}
Next, let $\currentDensity\left(\creaseAngle, \tv\right)$ denote the current density in the diffusion regime.
By continuity,
\begin{equation} \label{eq:current-continuity}
	\partialx{\prob}{\tv} + \partialx{\currentDensity}{\creaseAngle} = 0,
\end{equation}
which gives us a means to obtain the current density explicitly.

\begin{figure}
	\centering
	\includegraphics[width=0.65\linewidth]{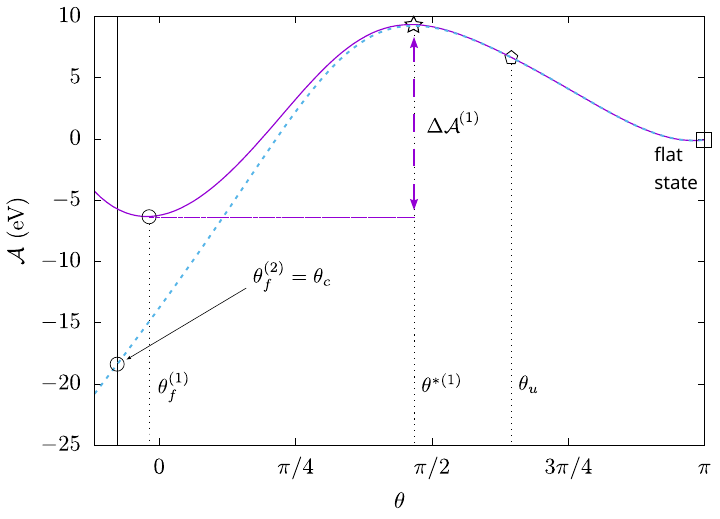}
	\caption{
		\capTitle{Example free energy landscapes for studying rate of state transitions.}
		$\creaseAngleFoldedMin^{(1)}$ and $\creaseAngleFoldedMin^{(2)}$ are the folded minima for the first and second example free energy landscapes, respectively.
		Note that, in contrast to $\creaseAngleFoldedMin^{(1)}$, $\creaseAngleFoldedMin^{(2)}$ is not a minima because the thermodynamic torque, $-\partialInline{\freeEnergy}{\creaseAngle}$, vanishes, but instead because the thermodynamic torque pins the crease at the boundary, $\creaseAngleClosed$.
		$\creaseAngleBarrier$ is the location of the free energy barrier and $\creaseAngleU$ is some crease angle just beyond the barrier.
		The transition rate is related to the current density at $\creaseAngleU$ relative to the probability mass about $\creaseAngleFoldedMin$.
	}
	\label{fig:energy-landscapes_escape}
\end{figure}

Let $\creaseAngleU$ denote some crease angle just beyond the free energy barrier, $\creaseAngleBarrier$, and $\creaseAngleFoldedMin$ denote the crease angle of the folded minima.
Examples are shown in \fref{fig:energy-landscapes_escape}.
Next we imagine an ensemble of molecular origami structures that are initialized about $\creaseAngleFoldedMin$ with energies that are less than $\freeEnergy\left(\creaseAngleStar\right) - \kB \T$.
The molecular origamis thermalize before unfolding and escaping to $\creaseAngle \rightarrow \pi$ where they are subsequently removed from the ensemble~\cite{balakrishnan2008elements,hanggi1990reaction}
Assume there exists a nontrivial equilibrium probability density, $\probEq$, and corresponding stationary current, $\currentDensityEq$.
The transition rate from folded to flat can be related to the stationary current by
\begin{equation} \label{eq:rate-ratio}
	\rate = \frac{\currentDensityEq}{\int_{\nbr{\creaseAngleFoldedMin}} \df{\creaseAngle} \: \probEq\left(\creaseAngle\right)},
\end{equation}
that is, the ratio of the flux at $\creaseAngleU$ to the probability mass around $\creaseAngleFoldedMin$, where $\nbr{\creaseAngleFoldedMin}$ is a neighborhood about $\creaseAngleFoldedMin$.
Recalling \eqref{eq:FPE} and \eqref{eq:current-continuity}, and integrating, we obtain~\cite{balakrishnan2008elements}
\begin{equation} \label{eq:rate0}
	\currentDensityEq = \frac{\kB \T}{\Imom \drag} \frac{\left(\probEq\left(\creaseAngle_f\right)\exp\left(\frac{\freeEnergy\left(\creaseAngle_f\right)}{\kB \T}\right) - \probEq\left(\creaseAngleU\right)\exp\left(\frac{\freeEnergy\left(\creaseAngleU\right)}{\kB \T}\right)\right)}{\int_{\creaseAngleFoldedMin}^{\creaseAngleBarrier} \df{\creaseAngle} \: \exp\left(\frac{\freeEnergy\left(\creaseAngle\right)}{\kB \T}\right)}.
\end{equation}
It is assumed that $\probEq\left(\creaseAngleU\right)$ is vanishingly small, and so this term is dropped.
The denominator in equation \eqref{eq:rate0} is approximated using Laplace's method, resulting in:
\begin{equation} \label{eq:rate1}
	\currentDensityEq \approx  \frac{\probEq\left(\creaseAngleFoldedMin\right)}{\Imom \drag} \sqrt{\frac{ \kB \T\left|\freeEnergy''\left(\creaseAngleBarrier\right)\right|}{2 \pi}} \: \exp\left(-\frac{\dA}{\kB \T}\right),
\end{equation}
where \hl{$\freeEnergy' = \partial \freeEnergy / \partial \creaseAngle$, $\freeEnergy'' = \partial^2 \freeEnergy / \partial \creaseAngle^2$, etc., and} $\dA = \freeEnergy\left(\creaseAngleBarrier\right) - \freeEnergy\left(\creaseAngleFoldedMin\right)$.

We pause here to consider the forms of \eqref{eq:rate-ratio} and \eqref{eq:rate1}.
Even if not much is known about the energy landscape beyond $\dA$, it is still evident that an Arrhenius type relationship should hold for the unfolding rate.
One of the simplest approximations that can be made regarding the transition rate of an energy barrier ``hopping'' process is given by
\begin{equation} \label{eq:tst}
	\rateTST = \frac{\kB \T}{h} \exp\left(-\frac{\dA}{\kB \T}\right),
\end{equation}
where $h$ is Planck's constant and $\kB \T / h$ provides a temperature dependent (inverse) time scale.
Equation \eqref{eq:tst} is a particular variant of \emph{transition state theory}.
There are many modifications and generalizations of transition state theory that are more or less well-suited for various cases.
See~\cite{hanggi1990reaction,vanden2005transition,bao2017variational} for example.

While the simplicity of \eqref{eq:tst} is attractive, further refinements can be made, due to the seminal work of Kramers~\cite{kramers1940brownian}, by returning to \eqref{eq:rate1} and including more information regarding the energy landscape.
By assumption, the system has thermalized in the energy well near the folded minima, $\creaseAngleFoldedMin$.
Then it is well known that, within this well, $\probEq\left(\creaseAngle\right) = C \exp\left(-\freeEnergy\left(\creaseAngle\right) / \kB \T\right)$ where $C$ is a normalization constant.
The normalization constant is difficult to evaluate; so instead, recognize that
\begin{equation}
	\probEq\left(\creaseAngle\right) = \probEq\left(\creaseAngleFoldedMin\right) \exp\left(\frac{\freeEnergy\left(\creaseAngleFoldedMin\right) - \freeEnergy\left(\creaseAngle\right)}{\kB \T}\right),
\end{equation}
and,
\begin{equation}
	\int_{\nbr{\creaseAngleFoldedMin}} \df{\creaseAngle} \: \probEq\left(\creaseAngle\right) = \probEq\left(\creaseAngleFoldedMin\right) \exp\left(\frac{\freeEnergy\left(\creaseAngleFoldedMin\right)}{\kB \T}\right) \int_{\nbr{\creaseAngleFoldedMin}} \df{\creaseAngle} \:  \exp\left(-\frac{\freeEnergy\left(\creaseAngle\right)}{\kB \T}\right).
\end{equation}
Now consider three separate cases\footnote{An additional assumption in all of three cases is that $\freeEnergy'\left(\creaseAngle\right) = 0 \implies \freeEnergy''\left(\creaseAngle\right) \neq 0$. This assumption is often made implicitly when employing Laplace's method in statistical mechanics.}: \begin{inparaenum}[1)] \item when the folded minima, $\creaseAngleFoldedMin$, is in the interior of the interval $\left[\creaseAngleClosed, \pi / 2\right]$ (e.g. $\creaseAngleFoldedMin^{(1)}$ in \fref{fig:energy-landscapes_escape}), \item when $\creaseAngleFoldedMin = \creaseAngleClosed$ and $\freeEnergy'\left(\creaseAngleClosed\right) = 0$, and \item when $\creaseAngleFoldedMin = \creaseAngleClosed$ and $\freeEnergy'\left(\creaseAngleClosed\right) > 0$ (e.g. $\creaseAngleFoldedMin^{(2)}$ in \fref{fig:energy-landscapes_escape}). \end{inparaenum}
For the interior case, we once again utilize Laplace's method so that
\begin{align}
	\int_{\nbr{\creaseAngleFoldedMin}} \df{\creaseAngle} \: \probEq\left(\creaseAngle\right) &\approx \probEq\left(\creaseAngleFoldedMin\right) \sqrt{\frac{2 \pi \kB \T}{\freeEnergy''\left(\creaseAngleFoldedMin\right)}}, \\
	\rate &\approx \frac{\sqrt{\freeEnergy''\left(\creaseAngleFoldedMin\right) \left|\freeEnergy''\left(\creaseAngleBarrier\right)\right|}}{2 \pi \Imom \drag} \exp\left(-\frac{\dA}{\kB \T}\right)
\end{align}
For the boundary cases, we need to adjust the standard Kramers approach.
When $\creaseAngleFoldedMin = \creaseAngleClosed$ and $\freeEnergy'\left(\creaseAngleClosed\right) = 0$, we can again use Laplace's method but only extend the upper bound of integration to infinity.
In this case, the probability mass takes the same form but with a factor of $1/2$ such that
\begin{align}
	\int_{\nbr{\creaseAngleFoldedMin}} \df{\creaseAngle} \: \probEq\left(\creaseAngle\right) &\approx \frac{1}{2} \probEq\left(\creaseAngleFoldedMin\right) \sqrt{\frac{2 \pi \kB \T}{\freeEnergy''\left(\creaseAngleFoldedMin\right)}}, \\
	\rate &\approx \frac{\sqrt{\freeEnergy''\left(\creaseAngleFoldedMin\right) \left|\freeEnergy''\left(\creaseAngleBarrier\right)\right|}}{\pi \Imom \drag} \exp\left(-\frac{\dA}{\kB \T}\right)
\end{align}
Last, consider the case when $\creaseAngleFoldedMin = \creaseAngleClosed$ and $\freeEnergy'\left(\creaseAngleClosed\right) > 0$.
Laplace's method is not applicable here, in general, as there is no guarantee that $\freeEnergy''\left(\creaseAngleFoldedMin\right) > 0$. 
We instead use the Taylor expansion $\freeEnergy\left(\creaseAngle\right) = \freeEnergy\left(\creaseAngleFoldedMin\right) + \freeEnergy'\left(\creaseAngleFoldedMin\right) \left(\creaseAngle - \creaseAngleFoldedMin\right)$ and obtain
\begin{align}
	\int_{\creaseAngleFoldedMin}^{\creaseAngle'} \df{\creaseAngle} \: \probEq\left(\creaseAngle\right) &\approx \probEq\left(\creaseAngleFoldedMin\right) \frac{\kB \T}{\freeEnergy'\left(\creaseAngleFoldedMin\right)}, \\
	\rate &\approx \frac{\freeEnergy'\left(\creaseAngleFoldedMin\right)}{\Imom \drag} \sqrt{\frac{\left|\freeEnergy''\left(\creaseAngleBarrier\right)\right|}{2 \pi \kB \T}} \exp\left(-\frac{\dA}{\kB \T}\right),
\end{align}
where $\creaseAngleClosed < \creaseAngle' \leq \creaseAngleBarrier$ and $\exp\left(-\freeEnergy\left(\creaseAngle'\right) / \kB \T\right)$ is dropped because it is assumed to be negligible.
Note that $\freeEnergy'\left(\creaseAngleFoldedMin\right) > 0$, as desired, if $\creaseAngleFoldedMin = \creaseAngleClosed$ is a local minima.
It can be seen in \fref{fig:energy-landscapes_Lu-and-d0}-\ref{fig:energy-landscapes_W-and-mu} that the free energy is--to a good approximation--linear in the neighborhood of the boundary minima for the systems considered therein.
Finally, the modified Kramers' rate is given by
\begin{equation} \label{eq:Kr}
	\rateKr = \begin{cases}
		\frac{\sqrt{\freeEnergy''\left(\creaseAngleFoldedMin\right) \left|\freeEnergy''\left(\creaseAngleBarrier\right)\right|}}{2 \pi \Imom \drag} \: \exp\left(-\frac{\dA}{\kB \T}\right), &  \creaseAngleFoldedMin \in (\creaseAngleClosed, \pi / 2] \\
		\frac{\sqrt{\freeEnergy''\left(\creaseAngleFoldedMin\right) \left|\freeEnergy''\left(\creaseAngleBarrier\right)\right|}}{\pi \Imom \drag} \: \exp\left(-\frac{\dA}{\kB \T}\right), &  \creaseAngleFoldedMin = \creaseAngleClosed \text{ and } \freeEnergy'\left(\creaseAngle\right) = 0 \\
		\frac{\freeEnergy'\left(\creaseAngleFoldedMin\right)}{\Imom \drag} \sqrt{\frac{\left|\freeEnergy''\left(\creaseAngleBarrier\right)\right|}{2 \pi \kB \T}} \: \exp\left(-\frac{\dA}{\kB \T}\right), & \text{otherwise}
	\end{cases}.
\end{equation}

\begin{figure}
	\centering
	\includegraphics[width=\linewidth]{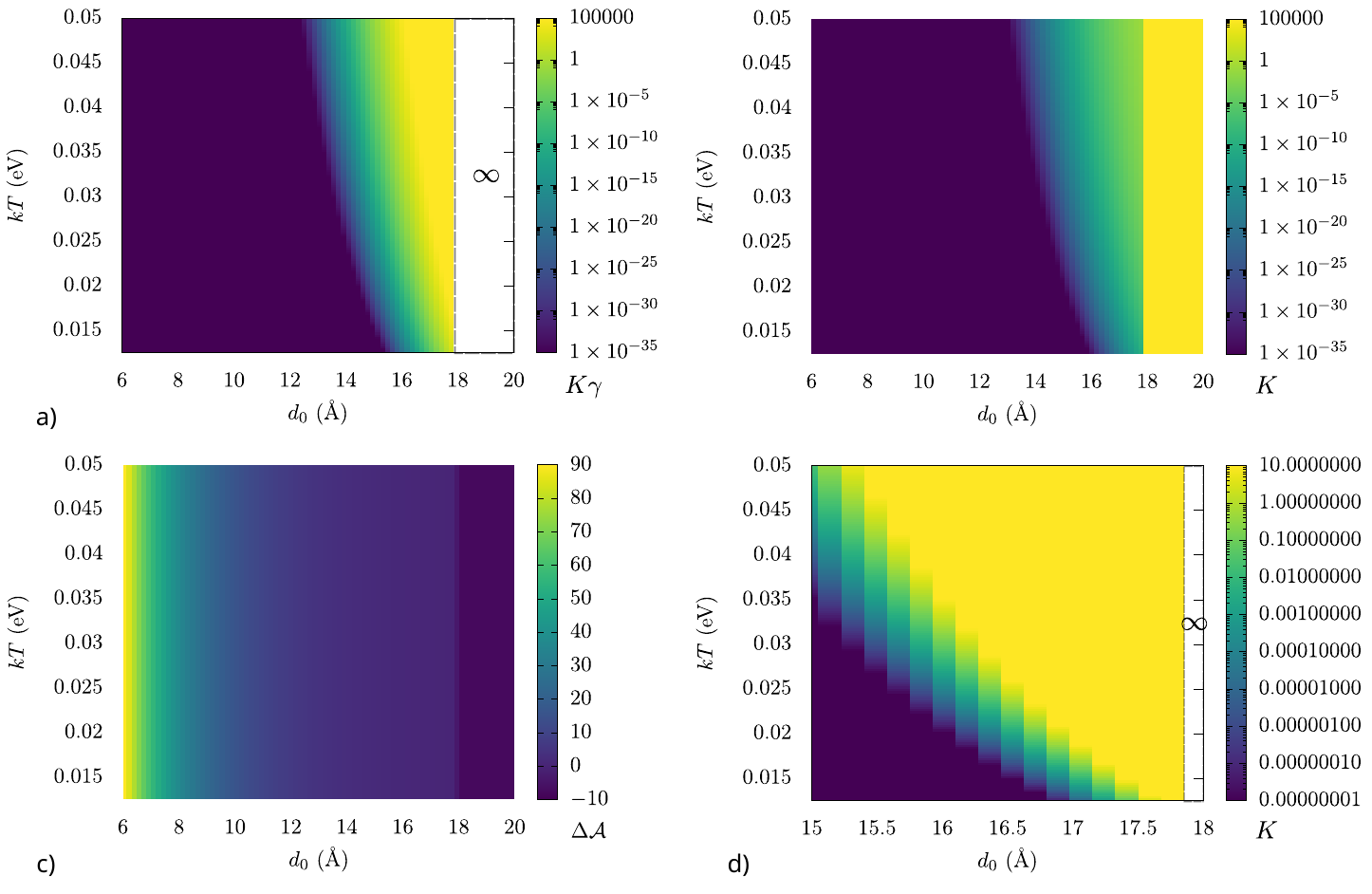}
	\caption{
		\capTitle{Interplay of temperature and crease diameter on the long-time, nonequilibrium behavior.}
		a) the transition rate approximated via a Kramers-like formula, $\rateKr \drag$, b) the rate approximated via transition state theory, $\rateTST$, c) the well depth, $\dA$, and d) the maximum of the two rate approximations, focused on the boundary of temporal stability.
	}
	\label{fig:rates_d0-vs-kT}
\end{figure}

\begin{figure}
	\centering
	\includegraphics[width=\linewidth]{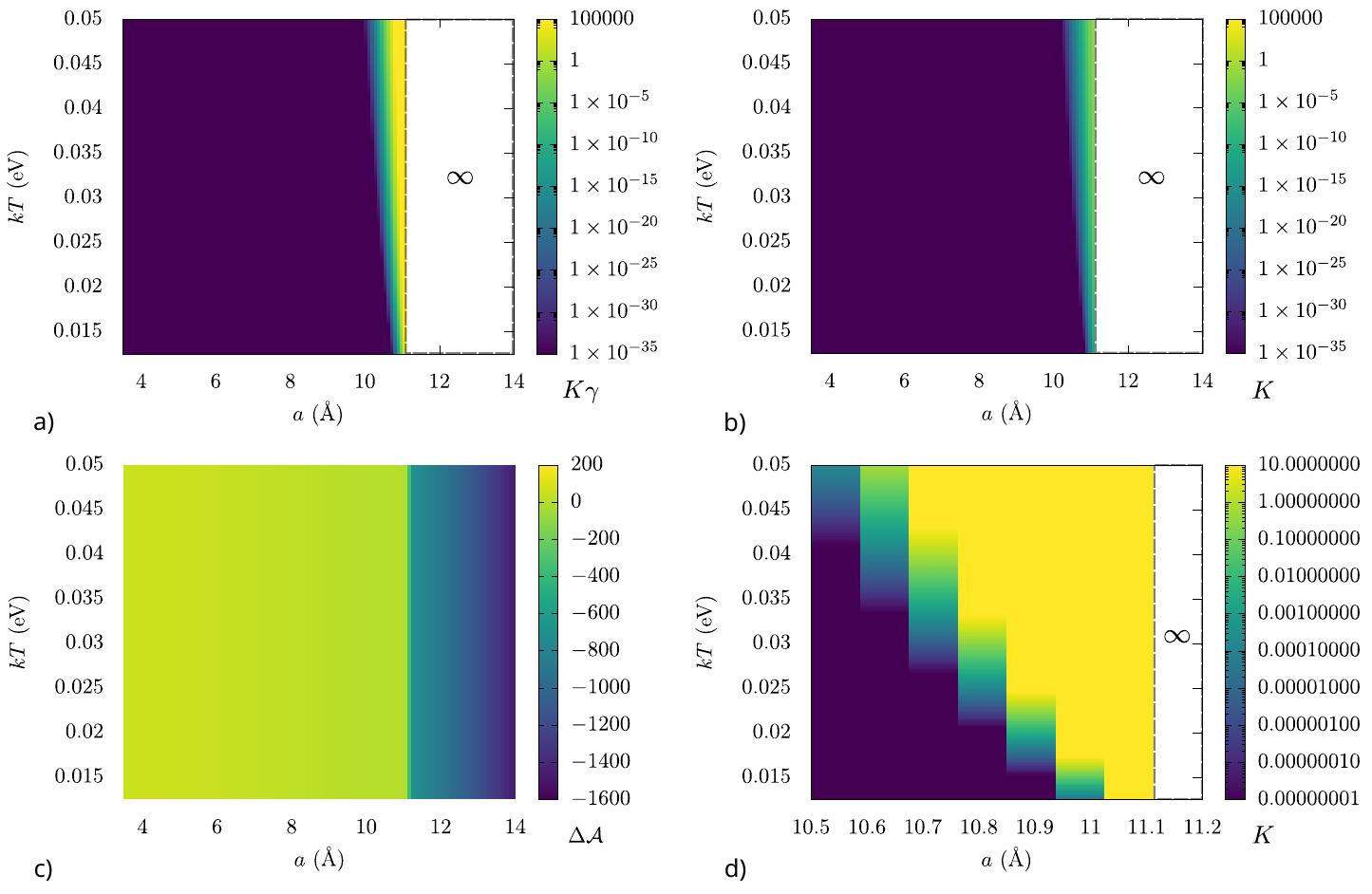}
	\caption{
		\capTitle{Interplay of temperature and membrane thickness on the long-time, nonequilibrium behavior.}
		a) the transition rate approximated via a Kramers-like formula, $\rateKr \drag$, b) the rate approximated via transition state theory, $\rateTST$, c) the well depth, $\dA$, and d) the maximum of the two rate approximations, focused on the boundary of temporal stability.
	}
	\label{fig:rates_a-vs-kT}
\end{figure}

\begin{figure}
	\centering
	\includegraphics[width=\linewidth]{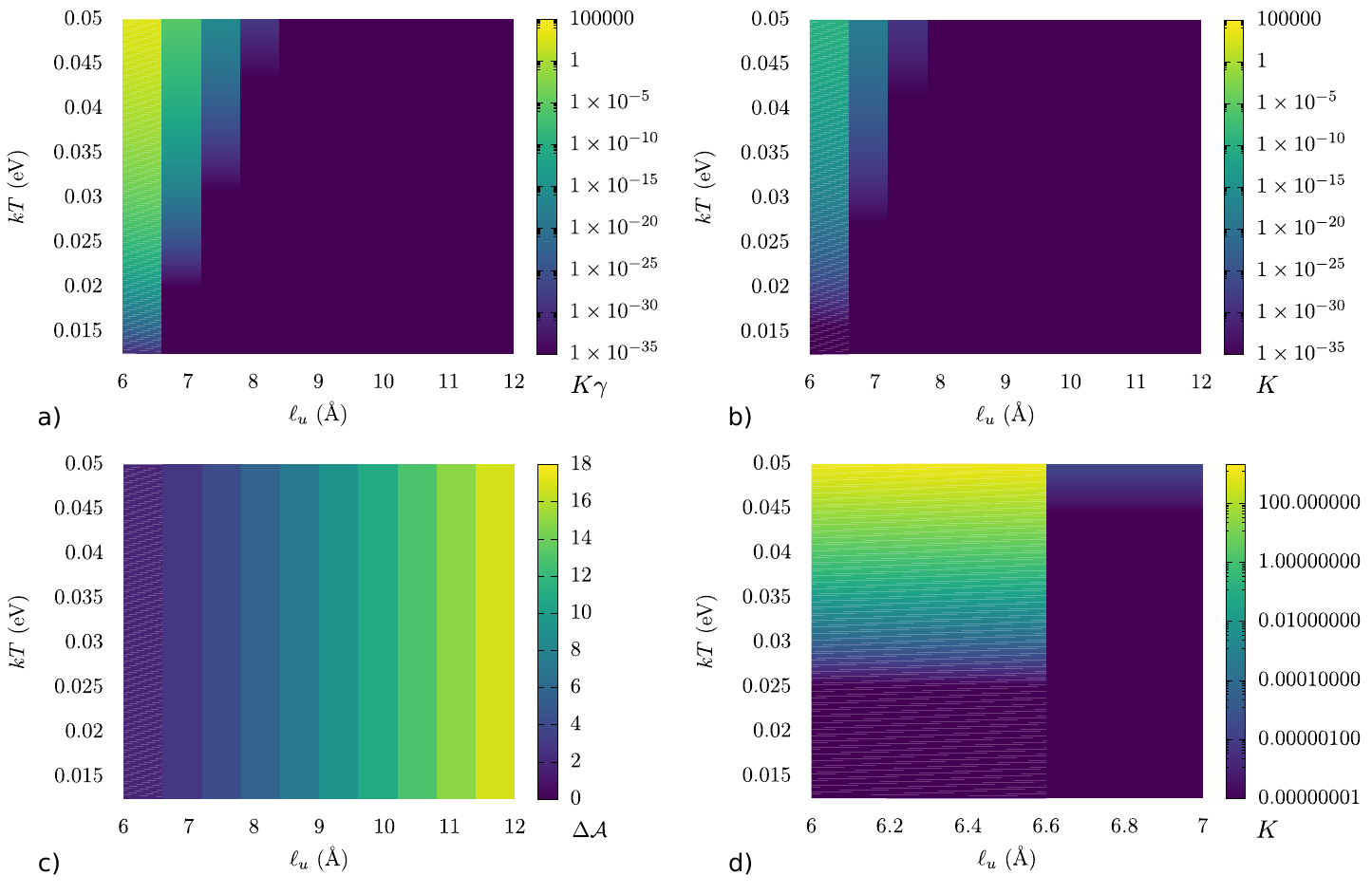}
	\caption{
		\capTitle{Interplay of temperature and membrane length, $\Lu$, on the long-time, nonequilibrium behavior.}
		a) the transition rate approximated via a Kramers-like formula, $\rateKr \drag$, b) the rate approximated via transition state theory, $\rateTST$, c) the well depth, $\dA$, and d) the maximum of the two rate approximations, focused on the boundary of temporal stability.
	}
	\label{fig:rates_Lu-vs-kT}
\end{figure}

With approximations for the transition rate from folded to flat in hand, we return to the graphene system considered in \fref{sec:grapheneEq}.
Recall that, for the base case, $\Lu = 24$ \angs, $\Lv = 70$ \angs, $\DO = 7$ \angs, $\lt = 3.5$ \angs, $\kbend = 0.95$ \eV, $\CvdW = 100$ \eV, and $\stericFactor = 1/6$.
Further, we assume the areal density is $\density = 0.763 \: \mathrm{mg} / \mathrm{m}^2$.
Then the moment of inertia (of the top membrane) about the center line of the crease is
\begin{equation} \label{eq:Imom}
	\Imom = \density \area \left(\frac{\Lu^2}{3} + \Lu \DO + \DO^2\right).
\end{equation}
In \fref{fig:rates_d0-vs-kT}-\ref{fig:rates_Lu-vs-kT}, we consider the interplay of temperature and geometry on the transition rates from flat to folded, i.e. we approximate the temporal stability of the folded energy well.
\Fref{fig:rates_d0-vs-kT} shows the transition rate properties of the system as a function of $\kB \T$ and $\DO$: \begin{inparaenum}[a)] \item heat map of the transition rate approximated via a Kramers-like formula \eqref{eq:Kr}, $\rateKr \drag$, \item the rate approximated via the simplest form of transition state theory \eqref{eq:tst}, $\rateTST$, \item the well depth, $\dA$, and \item the maximum of the two rate approximations (assuming $\drag$ as unity), $\max\left\{\rateKr \rateTST\right\}$, restricted to parameter space where the most significant changes in transition rates are occurring. \end{inparaenum}
To be precise, let $\creaseAngleFoldedMin$ and $\creaseAngleBarrier$ be the crease angles with the minimum and maximum free energies on the interval $\left[\creaseAngleClosed, \pi / 2\right]$, respectively.
Then, we let
\begin{equation}
	\dA =
	\begin{cases}
		\freeEnergy\left(\creaseAngleBarrier\right) - \freeEnergy\left(\creaseAngleFoldedMin\right), & \creaseAngleFoldedMin < \creaseAngleBarrier \\
		\freeEnergy\left(\creaseAngleFoldedMin\right) - \freeEnergy\left(\creaseAngleBarrier\right), & \text{otherwise}
	\end{cases}
\end{equation}
such that $\dA$ can be negative.
The origami is multistable provided $\dA > 0$.
Note that, although the molecular origami is multistable when $\DO \lessapprox 18$ \angs, the rate of unfolding becomes nonnegligible at smaller diameters, e.g. in the range of $12$ \angs $< \DO < 18$ \angs, depending on the temperature.
Here the role of temperature is two-fold: \begin{inparaenum}[1)] \item it increases the entropic repulsion between the two membranes which can alter the energy barrier, $\dA$, and \item the transition rate, for a fixed $\dA$, increases exponentially with temperature. \end{inparaenum}

\Fref{fig:rates_a-vs-kT} shows the transition rate properties as a function of $\kB \T$ and $\lt$.
As before, we assume that changes in thickness correspond with changes in bending stiffness such that $\lt \rightarrow \zeta \lt \implies \kbend \rightarrow \zeta^3 \kbend$.
Here, in contrast to varying $\DO$, one sees that generally systems which are multistable, i.e. $\dA > 0$, are also temporally stable.
There is only a narrow region, on the order of fractions of \angs, over which there is a distinction between equilibrium and temporal stability.
Increasing the thickness of the graphene increases the van der Waals attraction, while simultaneously making the crease more stiff and reducing entropic torque.
Recall \fref{fig:energy-landscapes_kT-and-thickness}.b.
As the thickness increases, the energy barrier moves up and to the left until the system is monostable; as this bifurcation occurs, the well depth is nonnegligible until $\creaseAngleBarrier \rightarrow \creaseAngleClosed$, which occurs over a small range, $10.6$ \angs $\lessapprox \lt \lessapprox 11.1$ \angs.

Lastly, \fref{fig:rates_Lu-vs-kT} shows the transition rate properties as a function of $\kB \T$ and $\Lu$.
As before, we see that the transition rate becomes nonnegligible prior to loss of equilibrium multistability, with a clear dependence on temperature.
Although the origami is multistable provided $\Lu \gtrapprox 6$ \angs, the transition rate begins to increase around $9$ \angs for higher temperatures (i.e. $0.05$ \eV) and $7$ \angs for lower temperatures ($0.015$ \eV).

\section{Closure} \label{sec:conclusion}

We address arguably the simplest possible problem that may be defined for a nanoscale origami structure---a folded elastic sheet with a crease. This deceptively ``simple" problem displays a rich interplay between several interesting elements such as the entropic repulsive force, the non-equilibrium nature of the unfolding process and mechanics. Using physically motivated simplifying assumptions we are able to make enough progress to understand the key factors that dictate the stability of a nanoscale origami structure under thermal fluctuations.\\

Future work can proceed several directions e.g. structures with multiple creases or multiple vertices. With connection to electromechanical coupling, applications for energy harvesting may be envisioned \cite{liu2013flexoelectricity, deng2014nanoscale}. The applications at the intersection of biophysics and biomedicine seem to be plentiful e.g. active matter \cite{kulkarni2023fluctuations}, DNA origami, among others. 

\begin{acknowledgments}
  MG acknowledges the support of the Air Force Research Laboratory. PS acknowledges support from the Cullen Professorship.
\end{acknowledgments}

\appendix

\section{Alternative approach to constraint at the crease} \label{app:constraint2}
Here the constraint $\h\left(0, \V\right) = 0$ is enforced via an energetic penalty\footnote{For a discussion of some of the nuances related to enforcing constraints exactly versus penalizing deviations from constraints via an energy penalty, see~\cite{weiner2012statistical}}.
The partition function is
\begin{equation} \label{eq:part-alt}
	\partFunc = \exp\left(-\frac{\Hcrease + \HvdW}{\kB \T}\right) \int \exp\left(-\frac{\Hbend\left[\h\right] + \HGauss\left[\h\right] + \frac{1}{2} \vPot \intOverSurf{\left(\h \cos \creaseAngle\right)^2} + \frac{1}{2} \pPot \int_0^{\Lv} \df{\V} \: \h^2 \left(0, \V\right)}{\kB \T}\right)
\end{equation}
where $\pPot$ is the magnitude of the penalty.
We will be interested in the limit of $\pPot \rightarrow \infty$.
The partition function is evaluated as
\begin{equation} \label{eq:part-alt-eval}
    \partFunc = \exp\left(-\frac{\Hcrease + \HvdW}{\kB \T}\right) \prod_{\left(m, n\right) \in \modeNums} \left(\frac{8 \pi \kB \T}{\area \left(\vPot \cos^2 \creaseAngle + \kbend\left(\qm^2 + \pn^2\right)\right)}\right) \left(\frac{8 \pi \kB \T}{\area \left(\frac{2 \pPot}{\Lu} + \vPot \cos^2 \creaseAngle + \kbend\left(\qm^2 + \pn^2\right)\right)}\right),
\end{equation}
Using \eqref{eq:hsqrd}, we have the result
\begin{equation}
    \avg{\h^2 \cos^2 \creaseAngle} = \frac{\kB \T \cos \creaseAngle}{16 \sqrt{\vPot \kbend}} + \frac{\kB \T \cos^2 \creaseAngle}{16 \sqrt{\left(\vPot\cos^2 \creaseAngle + \frac{2 \pPot}{\Lu}\right) \kbend}}.
\end{equation}
Note that, in the limit of $\pPot \rightarrow \infty$, this recovers the analogous result, \eqref{eq:hsqrdCons}, where the constraint was enforced exactly.
Again, let $\avg{\h^2 \cos^2 \creaseAngle} = \stericFactor \DAvg^2$.
Then, in the limit of $\pPot \rightarrow \infty$,
\begin{equation}
    \vPot = \frac{1}{\kbend} \left(\frac{\kB \T \cos \creaseAngle}{16 \stericFactor \DAvg^2}\right)^2.
\end{equation}

The free energy of the system can be obtained from \eqref{eq:free-energy}
\begin{equation} \label{eq:free-energy-fold2}
    \begin{split}
        \freeEnergy &= \Hcrease + \HvdW -\kB \T \sum_{\left(m, n\right) \in \modeNums} \log \frac{8 \pi \kB \T}{\area \left(\vPot \cos^2 \creaseAngle + \kbend\left(\qm^2 + \pn^2\right)\right)} -\kB \T \sum_{\left(m, n\right) \in \modeNums} \log \frac{8 \pi \kB \T}{\area \left(\frac{2 \pPot}{\Lu} + \vPot \cos^2 \creaseAngle + \kbend\left(\qm^2 + \pn^2\right)\right)}\\
        &= \Hcrease + \HvdW + \frac{\area}{256 \kbend \stericFactor} \left(\frac{\kB \T \cos \creaseAngle}{\DAvg}\right)^2 + \area \kB \T \sqrt{\frac{2 \pPot}{\Lu \kbend} + \left(\frac{\kB \T \cos^2 \creaseAngle}{256 \kbend \stericFactor \DAvg^2}\right)^2}\\
        &= \Hcrease + \HvdW + \frac{\area}{256 \kbend \stericFactor} \left(\frac{\kB \T \cos \creaseAngle}{\DAvg}\right)^2 + \area \kB \T \sqrt{\pPot}\left(\frac{1}{8\sqrt{2 \Lu \kbend}} + \frac{\sqrt{\Lu / \kbend^3} \smallpar}{8192\sqrt{2}} + \orderOf{\smallpar^2}\right),
    \end{split}
\end{equation}
where
\begin{equation}
    \smallpar = \left(\frac{\kB \T \cos^2 \creaseAngle}{256 \kbend \stericFactor \DAvg^2}\right)^2 / \pPot.
\end{equation}
Consider the last term of \eqref{eq:free-energy-fold2}.
We drop the first of the three terms in the parentheses as it represents a contribution to the free energy which is energetically inaccessible in the limit of $\pPot \rightarrow \infty$.
The last two terms vanish in the limit of $\pPot \rightarrow \infty$.
What remains is equivalent to the free energy obtained by enforcing the constraints exactly, \eqref{eq:free-energy-fold}.

\bibliographystyle{unsrtnat}
\bibliography{master}

\begin{thebibliography}{64}
\providecommand{\natexlab}[1]{#1}
\providecommand{\url}[1]{\texttt{#1}}
\expandafter\ifx\csname urlstyle\endcsname\relax
  \providecommand{\doi}[1]{doi: #1}\else
  \providecommand{\doi}{doi: \begingroup \urlstyle{rm}\Url}\fi

\bibitem[Boatti et~al.(2017)Boatti, Vasios, and Bertoldi]{boatti2017origami}
Elisa Boatti, Nikolaos Vasios, and Katia Bertoldi.
\newblock Origami metamaterials for tunable thermal expansion.
\newblock \emph{Advanced Materials}, 29\penalty0 (26):\penalty0 1700360, 2017.

\bibitem[Liu et~al.(2022)Liu, Pratapa, Misseroni, Tachi, and
  Paulino]{liu2022triclinic}
Ke~Liu, Phanisri~P Pratapa, Diego Misseroni, Tomohiro Tachi, and Glaucio~H
  Paulino.
\newblock Triclinic metamaterials by tristable origami with reprogrammable
  frustration.
\newblock \emph{Advanced Materials}, 34\penalty0 (43):\penalty0 2107998, 2022.

\bibitem[Miyazawa et~al.(2021)Miyazawa, Yasuda, Kim, Lynch, Tsujikawa,
  Kunimine, Raney, and Yang]{miyazawa2021heterogeneous}
Yasuhiro Miyazawa, Hiromi Yasuda, Hyungkyu Kim, James~H Lynch, Kosei Tsujikawa,
  Takahiro Kunimine, Jordan~R Raney, and Jinkyu Yang.
\newblock Heterogeneous origami-architected materials with variable stiffness.
\newblock \emph{Communications Materials}, 2\penalty0 (1):\penalty0 1--7, 2021.

\bibitem[Misseroni et~al.(2022)Misseroni, Pratapa, Liu, and
  Paulino]{misseroni2022experimental}
Diego Misseroni, Phanisri~P Pratapa, Ke~Liu, and Glaucio~H Paulino.
\newblock Experimental realization of tunable poisson’s ratio in deployable
  origami metamaterials.
\newblock \emph{Extreme Mechanics Letters}, 53:\penalty0 101685, 2022.

\bibitem[Zhai et~al.(2020)Zhai, Wang, Lin, Wu, and Jiang]{zhai2020situ}
Zirui Zhai, Yong Wang, Ken Lin, Lingling Wu, and Hanqing Jiang.
\newblock In situ stiffness manipulation using elegant curved origami.
\newblock \emph{Science advances}, 6\penalty0 (47):\penalty0 eabe2000, 2020.

\bibitem[Silverberg et~al.(2014)Silverberg, Evans, McLeod, Hayward, Hull,
  Santangelo, and Cohen]{silverberg2014using}
Jesse~L Silverberg, Arthur~A Evans, Lauren McLeod, Ryan~C Hayward, Thomas Hull,
  Christian~D Santangelo, and Itai Cohen.
\newblock Using origami design principles to fold reprogrammable mechanical
  metamaterials.
\newblock \emph{science}, 345\penalty0 (6197):\penalty0 647--650, 2014.

\bibitem[Schenk and Guest(2013)]{schenk2013geometry}
Mark Schenk and Simon~D Guest.
\newblock Geometry of miura-folded metamaterials.
\newblock \emph{Proceedings of the National Academy of Sciences}, 110\penalty0
  (9):\penalty0 3276--3281, 2013.

\bibitem[Liu et~al.(2018)Liu, Silverberg, Evans, Santangelo, Lang, Hull, and
  Cohen]{liu2018topological}
Bin Liu, Jesse~L Silverberg, Arthur~A Evans, Christian~D Santangelo, Robert~J
  Lang, Thomas~C Hull, and Itai Cohen.
\newblock Topological kinematics of origami metamaterials.
\newblock \emph{Nature Physics}, 14\penalty0 (8):\penalty0 811--815, 2018.

\bibitem[Grasinger et~al.(2022)Grasinger, Gillman, and
  Buskohl]{grasinger2022multistability}
Matthew Grasinger, Andrew Gillman, and Philip~R Buskohl.
\newblock Multistability, symmetry and geometric conservation in eightfold
  waterbomb origami.
\newblock \emph{Proceedings of the Royal Society A}, 478\penalty0
  (2268):\penalty0 20220270, 2022.

\bibitem[Pratapa et~al.(2019)Pratapa, Liu, and Paulino]{pratapa2019geometric}
Phanisri~P Pratapa, Ke~Liu, and Glaucio~H Paulino.
\newblock Geometric mechanics of origami patterns exhibiting poisson’s ratio
  switch by breaking mountain and valley assignment.
\newblock \emph{Physical Review Letters}, 122\penalty0 (15):\penalty0 155501,
  2019.

\bibitem[Brunck et~al.(2016)Brunck, Lechenault, Reid, and
  Adda-Bedia]{brunck2016elastic}
V~Brunck, F~Lechenault, A~Reid, and M~Adda-Bedia.
\newblock Elastic theory of origami-based metamaterials.
\newblock \emph{Physical Review E}, 93\penalty0 (3):\penalty0 033005, 2016.

\bibitem[Sessions et~al.(2019)Sessions, Cook, Fuchi, Gillman, Huff, and
  Buskohl]{sessions2019origami}
Deanna Sessions, Alexander Cook, Kazuko Fuchi, Andrew Gillman, Gregory Huff,
  and Philip Buskohl.
\newblock Origami-inspired frequency selective surface with fixed frequency
  response under folding.
\newblock \emph{Sensors}, 19\penalty0 (21):\penalty0 4808, 2019.

\bibitem[Novelino et~al.(2020)Novelino, Ze, Wu, Paulino, and
  Zhao]{novelino2020untethered}
Larissa~S Novelino, Qiji Ze, Shuai Wu, Glaucio~H Paulino, and Ruike Zhao.
\newblock Untethered control of functional origami microrobots with distributed
  actuation.
\newblock \emph{Proceedings of the National Academy of Sciences}, 117\penalty0
  (39):\penalty0 24096--24101, 2020.

\bibitem[Wu et~al.(2021)Wu, Ze, Dai, Udipi, Paulino, and
  Zhao]{wu2021stretchable}
Shuai Wu, Qiji Ze, Jize Dai, Nupur Udipi, Glaucio~H Paulino, and Ruike Zhao.
\newblock Stretchable origami robotic arm with omnidirectional bending and
  twisting.
\newblock \emph{Proceedings of the National Academy of Sciences}, 118\penalty0
  (36), 2021.

\bibitem[Fang et~al.(2017)Fang, Zhang, and Wang]{fang2017origami}
Hongbin Fang, Yetong Zhang, and KW~Wang.
\newblock Origami-based earthworm-like locomotion robots.
\newblock \emph{Bioinspiration \& biomimetics}, 12\penalty0 (6):\penalty0
  065003, 2017.

\bibitem[Treml et~al.(2018)Treml, Gillman, Buskohl, and Vaia]{treml2018origami}
Benjamin Treml, Andrew Gillman, Philip Buskohl, and Richard Vaia.
\newblock Origami mechanologic.
\newblock \emph{Proceedings of the National Academy of Sciences}, 115\penalty0
  (27):\penalty0 6916--6921, 2018.

\bibitem[Kawasaki(1991)]{kawasaki1991relation}
Toshikazu Kawasaki.
\newblock On the relation between mountain-creases and valley-creases of a flat
  origami.
\newblock In \emph{Proceedings of the First International Meeting of Origami
  Science and Technology, 1991}, 1991.

\bibitem[Lang(2017)]{lang2017twists}
Robert~J Lang.
\newblock \emph{Twists, tilings, and tessellations: mathematical methods for
  geometric origami}.
\newblock CRC press, 2017.

\bibitem[Demaine and O'Rourke(2007)]{demaine2007geometric}
Erik~D Demaine and Joseph O'Rourke.
\newblock \emph{Geometric folding algorithms: linkages, origami, polyhedra}.
\newblock Cambridge university press, 2007.

\bibitem[belcastro and Hull(2002)]{hull2002modelling}
sarah-marie belcastro and Thomas~C Hull.
\newblock Modelling the folding of paper into three dimensions using affine
  transformations.
\newblock \emph{Linear Algebra and its applications}, 348\penalty0
  (1-3):\penalty0 273--282, 2002.

\bibitem[Jiang et~al.(2019)Jiang, Liu, Liu, Wang, and
  Ding]{jiang2019rationally}
Qiao Jiang, Shaoli Liu, Jianbing Liu, Zhen-Gang Wang, and Baoquan Ding.
\newblock Rationally designed dna-origami nanomaterials for drug delivery in
  vivo.
\newblock \emph{Advanced Materials}, 31\penalty0 (45):\penalty0 1804785, 2019.

\bibitem[Udomprasert and Kangsamaksin(2017)]{udomprasert2017dna}
Anuttara Udomprasert and Thaned Kangsamaksin.
\newblock Dna origami applications in cancer therapy.
\newblock \emph{Cancer science}, 108\penalty0 (8):\penalty0 1535--1543, 2017.

\bibitem[Weiden and Bastings(2021)]{weiden2021dna}
Jorieke Weiden and Maartje~MC Bastings.
\newblock Dna origami nanostructures for controlled therapeutic drug delivery.
\newblock \emph{Current Opinion in Colloid \& Interface Science}, 52:\penalty0
  101411, 2021.

\bibitem[Ge et~al.(2020)Ge, Guo, Wu, Li, Sun, Hou, Shi, Song, Wang, Fan,
  et~al.]{ge2020dna}
Zhilei Ge, Linjie Guo, Guangqi Wu, Jiang Li, Yunlong Sun, Yingqin Hou, Jiye
  Shi, Shiping Song, Lihua Wang, Chunhai Fan, et~al.
\newblock Dna origami-enabled engineering of ligand--drug conjugates for
  targeted drug delivery.
\newblock \emph{Small}, 16\penalty0 (16):\penalty0 1904857, 2020.

\bibitem[Chandrasekaran et~al.(2016)Chandrasekaran, Anderson, Kizer, Halvorsen,
  and Wang]{chandrasekaran2016beyond}
Arun~Richard Chandrasekaran, Nate Anderson, Megan Kizer, Ken Halvorsen, and
  Xing Wang.
\newblock Beyond the fold: Emerging biological applications of dna origami.
\newblock \emph{ChemBioChem}, 17\penalty0 (12):\penalty0 1081--1089, 2016.

\bibitem[Liu et~al.(2023)Liu, Loh, Siti, Too, Anderson, and Wang]{liu2023light}
Xiao~Rui Liu, Iong~Ying Loh, Winna Siti, Hon~Lin Too, Tommy Anderson, and
  Zhisong Wang.
\newblock A light-operated integrated dna walker--origami system beyond bridge
  burning.
\newblock \emph{Nanoscale Horizons}, 8\penalty0 (6):\penalty0 827--841, 2023.

\bibitem[Cho et~al.(2011)Cho, Keung, Verellen, Lagae, Moshchalkov, Van~Dorpe,
  and Gracias]{cho2011nanoscale}
Jeong-Hyun Cho, Michael~D Keung, Niels Verellen, Liesbet Lagae, Victor
  Moshchalkov, Pol Van~Dorpe, and David~H Gracias.
\newblock Nanoscale origami for 3d optics.
\newblock \emph{Small}, 7\penalty0 (14):\penalty0 1943--1948, 2011.

\bibitem[Zhao et~al.(2022)Zhao, Zhang, Chen, Yang, and
  Kitipornchai]{zhao2022enhanced}
Shaoyu Zhao, Yingyan Zhang, Da~Chen, Jie Yang, and Sritawat Kitipornchai.
\newblock Enhanced thermal buckling resistance of folded graphene reinforced
  nanocomposites with negative thermal expansion: From atomistic study to
  continuum mechanics modelling.
\newblock \emph{Composite Structures}, 279:\penalty0 114872, 2022.

\bibitem[Zhao et~al.(2021)Zhao, Zhang, Yang, and
  Kitipornchai]{zhao2021significantly}
Shaoyu Zhao, Yingyan Zhang, Jie Yang, and Sritawat Kitipornchai.
\newblock Significantly improved interfacial shear strength in graphene/copper
  nanocomposite via wrinkles and functionalization: A molecular dynamics study.
\newblock \emph{Carbon}, 174:\penalty0 335--344, 2021.

\bibitem[Ahmadpoor et~al.(2017)Ahmadpoor, Wang, Huang, and
  Sharma]{ahmadpoor2017thermal}
Fatemeh Ahmadpoor, Peng Wang, Rui Huang, and Pradeep Sharma.
\newblock Thermal fluctuations and effective bending stiffness of elastic thin
  sheets and graphene: A nonlinear analysis.
\newblock \emph{Journal of the Mechanics and Physics of Solids}, 107:\penalty0
  294--319, 2017.

\bibitem[Fasolino et~al.(2007)Fasolino, Los, and
  Katsnelson]{fasolino2007intrinsic}
Annalisa Fasolino, JH~Los, and Mikhail~I Katsnelson.
\newblock Intrinsic ripples in graphene.
\newblock \emph{Nature materials}, 6\penalty0 (11):\penalty0 858--861, 2007.

\bibitem[Thibado et~al.(2020)Thibado, Kumar, Singh, Ruiz-Garcia, Lasanta, and
  Bonilla]{thibado2020fluctuation}
PM~Thibado, P~Kumar, Surendra Singh, M~Ruiz-Garcia, A~Lasanta, and LL~Bonilla.
\newblock Fluctuation-induced current from freestanding graphene.
\newblock \emph{Physical Review E}, 102\penalty0 (4):\penalty0 042101, 2020.

\bibitem[De~Parga et~al.(2008)De~Parga, Calleja, Borca, Passeggi~Jr, Hinarejos,
  Guinea, and Miranda]{de2008periodically}
AL~V{\'a}zquez De~Parga, Fabi{\'a}n Calleja, BMCG Borca, MCG Passeggi~Jr,
  JJ~Hinarejos, Francisco Guinea, and Rodolfo Miranda.
\newblock Periodically rippled graphene: growth and spatially resolved
  electronic structure.
\newblock \emph{Physical review letters}, 100\penalty0 (5):\penalty0 056807,
  2008.

\bibitem[Helfrich(1978)]{helfrich1978steric}
Wolfgang Helfrich.
\newblock Steric interaction of fluid membranes in multilayer systems.
\newblock \emph{Zeitschrift f{\"u}r Naturforschung A}, 33\penalty0
  (3):\penalty0 305--315, 1978.

\bibitem[Helfrich and Servuss(1984)]{helfrich1984undulations}
Wolfgang Helfrich and R~M Servuss.
\newblock Undulations, steric interaction and cohesion of fluid membranes.
\newblock \emph{Il Nuovo Cimento D}, 3:\penalty0 137--151, 1984.

\bibitem[Hanlumyuang et~al.(2014)Hanlumyuang, Liu, and
  Sharma]{hanlumyuang2014revisiting}
Yuranan Hanlumyuang, Liping Liu, and Pradeep Sharma.
\newblock Revisiting the entropic force between fluctuating biological
  membranes.
\newblock \emph{Journal of the Mechanics and Physics of Solids}, 63:\penalty0
  179--186, 2014.

\bibitem[Freund(2013)]{freund2013entropic}
LB~Freund.
\newblock Entropic pressure between biomembranes in a periodic stack due to
  thermal fluctuations.
\newblock \emph{Proceedings of the National Academy of Sciences}, 110\penalty0
  (6):\penalty0 2047--2051, 2013.

\bibitem[Lu and Podgornik(2015)]{lu2015effective}
Bing-Sui Lu and Rudolf Podgornik.
\newblock Effective interactions between fluid membranes.
\newblock \emph{Physical Review E}, 92\penalty0 (2):\penalty0 022112, 2015.

\bibitem[Wennerstr{\"o}m and Olsson(2014)]{wennerstrom2014undulation}
H{\aa}kan Wennerstr{\"o}m and Ulf Olsson.
\newblock The undulation force; theoretical results versus experimental
  demonstrations.
\newblock \emph{Advances in colloid and interface science}, 208:\penalty0
  10--13, 2014.

\bibitem[Mozaffari et~al.(2021)Mozaffari, Ahmadpoor, and
  Sharma]{mozaffari2021flexoelectricity}
Kosar Mozaffari, Fatemeh Ahmadpoor, and Pradeep Sharma.
\newblock Flexoelectricity and the entropic force between fluctuating fluid
  membranes.
\newblock \emph{Mathematics and Mechanics of Solids}, 26\penalty0
  (12):\penalty0 1760--1778, 2021.

\bibitem[Liang and Purohit(2018)]{liang2018method}
Xiaojun Liang and Prashant~K Purohit.
\newblock A method to compute elastic and entropic interactions of membrane
  inclusions.
\newblock \emph{Extreme mechanics letters}, 18:\penalty0 29--35, 2018.

\bibitem[Ahmadpoor et~al.(2022)Ahmadpoor, Zou, and Gao]{ahmadpoor2022entropic}
Fatemeh Ahmadpoor, Guijin Zou, and Huajian Gao.
\newblock Entropic interactions of 2d materials with cellular membranes:
  Parallel versus perpendicular approaching modes.
\newblock \emph{Mechanics of Materials}, 174:\penalty0 104414, 2022.

\bibitem[Zhu et~al.(2022)Zhu, Leng, Jiang, Chang, Zhang, and
  Gao]{zhu2022thermal}
Fangyan Zhu, Jiantao Leng, Jin-Wu Jiang, Tienchong Chang, Tongyi Zhang, and
  Huajian Gao.
\newblock Thermal-fluctuation gradient induced tangential entropic forces in
  layered two-dimensional materials.
\newblock \emph{Journal of the Mechanics and Physics of Solids}, 163:\penalty0
  104871, 2022.

\bibitem[Chen and Kulkarni(2015)]{chen2015entropic}
Dengke Chen and Yashashree Kulkarni.
\newblock Entropic interaction between fluctuating twin boundaries.
\newblock \emph{Journal of the Mechanics and Physics of Solids}, 84:\penalty0
  59--71, 2015.

\bibitem[Mannattil et~al.(2022)Mannattil, Schwarz, and
  Santangelo]{mannattil2022thermal}
Manu Mannattil, Jennifer~M Schwarz, and Christian~D Santangelo.
\newblock Thermal fluctuations of singular bar-joint mechanisms.
\newblock \emph{Physical Review Letters}, 128\penalty0 (20):\penalty0 208005,
  2022.

\bibitem[Rocklin et~al.(2018)Rocklin, Vitelli, and Mao]{rocklin2018folding}
D~Rocklin, Vincenzo Vitelli, and Xiaoming Mao.
\newblock Folding mechanisms at finite temperature.
\newblock \emph{arXiv preprint arXiv:1802.02704}, 2018.

\bibitem[Yong and Mahadevan(2014)]{yong2014statistical}
Ee~Hou Yong and Lakshminarayanan Mahadevan.
\newblock Statistical mechanics and shape transitions in microscopic plates.
\newblock \emph{Physical Review Letters}, 112\penalty0 (4):\penalty0 048101,
  2014.

\bibitem[Yang et~al.(2021)Yang, Zhang, Hu, Penev, and
  Yakobson]{yang2021energetics}
Yang Yang, Zhuhua Zhang, Zhili Hu, Evgeni~S Penev, and Boris~I Yakobson.
\newblock Energetics of graphene origami and their ``spatial resolution''.
\newblock \emph{MRS Bulletin}, pages 1--6, 2021.

\bibitem[Kulkarni(2023)]{kulkarni2023fluctuations}
Yashashree Kulkarni.
\newblock Fluctuations of active membranes with nonlinear curvature elasticity.
\newblock \emph{Journal of the Mechanics and Physics of Solids}, 173:\penalty0
  105240, 2023.

\bibitem[Chen and Santangelo(2018)]{chen2018branches}
Bryan Gin-ge Chen and Christian~D Santangelo.
\newblock Branches of triangulated origami near the unfolded state.
\newblock \emph{Physical Review X}, 8\penalty0 (1):\penalty0 011034, 2018.

\bibitem[Zhou et~al.(2023)Zhou, Grasinger, Buskohl, and
  Bhattacharya]{zhou2023low}
Hao Zhou, Matthew Grasinger, Philip Buskohl, and Kaushik Bhattacharya.
\newblock Low energy fold paths in multistable origami structures.
\newblock \emph{International Journal of Solids and Structures}, page 112125,
  2023.

\bibitem[Tadmor(2001)]{tadmor2001london}
Rafael Tadmor.
\newblock The london-van der waals interaction energy between objects of
  various geometries.
\newblock \emph{Journal of physics: Condensed matter}, 13\penalty0
  (9):\penalty0 L195, 2001.

\bibitem[Helfrich(1973)]{helfrich1973elastic}
Wolfgang Helfrich.
\newblock Elastic properties of lipid bilayers: theory and possible
  experiments.
\newblock \emph{Zeitschrift f{\"u}r Naturforschung c}, 28\penalty0
  (11-12):\penalty0 693--703, 1973.

\bibitem[Fredrickson(2006)]{fredrickson2006equilibrium}
Glenn Fredrickson.
\newblock \emph{The equilibrium theory of inhomogeneous polymers}.
\newblock Number 134. Oxford University Press on Demand, 2006.

\bibitem[Grasinger et~al.(2021)Grasinger, Mozaffari, and
  Sharma]{grasinger2021flexoelectricity}
Matthew Grasinger, Kosar Mozaffari, and Pradeep Sharma.
\newblock Flexoelectricity in soft elastomers and the molecular mechanisms
  underpinning the design and emergence of giant flexoelectricity.
\newblock \emph{Proceedings of the National Academy of Sciences}, 118\penalty0
  (21):\penalty0 e2102477118, 2021.

\bibitem[H{\"a}nggi et~al.(1990)H{\"a}nggi, Talkner, and
  Borkovec]{hanggi1990reaction}
Peter H{\"a}nggi, Peter Talkner, and Michal Borkovec.
\newblock Reaction-rate theory: fifty years after kramers.
\newblock \emph{Reviews of modern physics}, 62\penalty0 (2):\penalty0 251,
  1990.

\bibitem[Vanden-Eijnden and Tal(2005)]{vanden2005transition}
Eric Vanden-Eijnden and Fabio~A Tal.
\newblock Transition state theory: Variational formulation, dynamical
  corrections, and error estimates.
\newblock \emph{The Journal of chemical physics}, 123\penalty0 (18), 2005.

\bibitem[Balakrishnan(2008)]{balakrishnan2008elements}
Venkataraman Balakrishnan.
\newblock \emph{Elements of nonequilibrium statistical mechanics}, volume~3.
\newblock Springer, 2008.

\bibitem[Bao and Truhlar(2017)]{bao2017variational}
Junwei~Lucas Bao and Donald~G Truhlar.
\newblock Variational transition state theory: theoretical framework and recent
  developments.
\newblock \emph{Chemical Society Reviews}, 46\penalty0 (24):\penalty0
  7548--7596, 2017.

\bibitem[Leadbetter et~al.(2023)Leadbetter, Purohit, and
  Reina]{leadbetter2023statistical}
Travis Leadbetter, Prashant~K Purohit, and Celia Reina.
\newblock A statistical mechanics framework for constructing non-equilibrium
  thermodynamic models.
\newblock \emph{arXiv preprint arXiv:2309.07112}, 2023.

\bibitem[Kramers(1940)]{kramers1940brownian}
Hendrik~Anthony Kramers.
\newblock Brownian motion in a field of force and the diffusion model of
  chemical reactions.
\newblock \emph{Physica}, 7\penalty0 (4):\penalty0 284--304, 1940.

\bibitem[Liu and Sharma(2013)]{liu2013flexoelectricity}
LP~Liu and P~Sharma.
\newblock Flexoelectricity and thermal fluctuations of lipid bilayer membranes:
  Renormalization of flexoelectric, dielectric, and elastic properties.
\newblock \emph{Physical Review E}, 87\penalty0 (3):\penalty0 032715, 2013.

\bibitem[Deng et~al.(2014)Deng, Kammoun, Erturk, and Sharma]{deng2014nanoscale}
Qian Deng, Mejdi Kammoun, Alper Erturk, and Pradeep Sharma.
\newblock Nanoscale flexoelectric energy harvesting.
\newblock \emph{International Journal of Solids and Structures}, 51\penalty0
  (18):\penalty0 3218--3225, 2014.

\bibitem[Weiner(2012)]{weiner2012statistical}
Jerome~Harris Weiner.
\newblock \emph{Statistical mechanics of elasticity}.
\newblock Courier Corporation, 2012.

\end{thebibliography}

\end{document}